\documentclass[printer]{aa}

\usepackage{txfonts}
\usepackage{graphicx}
\usepackage{rotating}
\usepackage{natbib}
\usepackage{mathrsfs}
\bibpunct{(}{)}{;}{a}{}{,}

\begin{document}

\title{A molecular survey of outflow gas: velocity-dependent shock chemistry
and the peculiar composition of the EHV gas\thanks{Based on observations 
carried out with the IRAM 30m Telescope. 
IRAM is supported by INSU/CNRS (France), MPG (Germany) and IGN (Spain).}}

\titlerunning{A molecular survey of outflow gas}

\author{M. Tafalla \inst{1}
\and
J. Santiago-Garc\'{\i}a \inst{1,2}
\and 
A. Hacar \inst{1}
\and
R. Bachiller \inst{1}
}

\authorrunning{Tafalla et al.}

\institute{Observatorio Astron\'omico Nacional (IGN),
Alfonso XII 3, E-28014 Madrid,
Spain
\and
Instituto de Radioastronom\'{\i}a Milim\'etrica 
(IRAM), Avenida Divina Pastora 7, N\'ucleo Central,
E 18012 Granada, Spain
}

\offprints{M. Tafalla \email{m.tafalla@oan.es}}
\date{Received -- / Accepted -- }

\abstract
{Bipolar outflows from Class 0 protostars often present 
two components in their CO spectra that have different kinematic behaviors:
a smooth outflow wing and a discrete, extremely high-velocity (EHV)
peak.}
{To better understand the origin of these two outflow components,
we have studied and compared their molecular composition.}
{We carried out a molecular survey  
of the outflows powered by L1448-mm and
IRAS 04166+2706, two sources with prominent wing
and EHV components.
For each source, we observed a number of
molecular lines towards the brightest outflow position
and used them to determine column densities for 12
different molecular species.}
{The molecular composition of the two outflows is very
similar. It presents systematic changes with
velocity that we analyze by dividing the 
outflow in three chemical regimes, 
two of them associated with the wing component and the
other the EHV gas. 
The analysis of the two wing regimes 
shows that species like
H$_2$CO and CH$_3$OH favor the low-velocity gas, while
SiO and HCN are more abundant in the fastest gas.
This fastest wing gas presents strong similarities
with the composition of the ``chemically active'' L1157 outflow
(whose abundances we re-evaluate in an appendix).
We find that the EHV regime is relatively
rich in O-bearing species compared to the wing regime. 
The EHV gas is not only detected
in CO and SiO (already reported elsewhere), but 
also in SO, CH$_3$OH, and H$_2$CO (newly reported here),
with a tentative detection in HCO$^+$.
At the same time, the EHV regime is relatively poor in
C-bearing molecules like CS and HCN, for which we only
obtain weak detections or upper limits despite deep integrations.
We suggest that this difference in composition 
arises from a lower C/O ratio in the EHV gas.}
{The different chemical compositions of the wing and EHV 
regimes suggest that these two
outflow components have different physical origins.
The wing component is better explained by 
shocked ambient gas, although none of the existing shock models
explains all observed features.
We hypothesize that the peculiar composition of the EHV
gas reflects its origin as a dense
wind from the protostar or its surrounding disk.}


\maketitle

%

\section{Introduction}

Outflows from the youngest protostars often
display the simplest
geometry and the highest degree of symmetry, 
and they are expected to reflect more faithfully than
other outflows the properties of the still-mysterious 
driving wind (e.g., \citealt{bac99}).
Among the youngest outflows known, one group
stands out for its pristine appearance in maps 
and the presence in their CO spectra of distinct
extremely high velocity (EHV) features. 
These features are secondary emission peaks at
the highest outflow velocities that tend to appear
symmetrically placed from the driving source
along the outflow axis. 
The prototype of this group of outflows is the one
driven by the 9~L$_\odot$ Class 0 protostar L1448-mm, 
located in the Perseus molecular cloud and
first identified by \citet{bac90}.
Molecular observations of the L1448 outflow
have shown that the moving material 
in this system 
lies in two different components: a
pair of conical shells and a highly 
collimated jet. In the spectra, the
emission from the shells appears 
in the form of a gradual outflow wing, while the
emission from the jet arises from the faster EHV
feature \citep{bac95,joe07,mau10,hir10}.

While the shell of relatively slow gas in the L1448 outflow
is similar to what is seen in more evolved outflows 
like L1551  (e.g., \citealt{mor87}), and therefore likely composed 
of accelerated ambient gas,  
the nature of the EHV component is still mysterious.
The symmetry, high speed, and fragmented
appearance of the EHV gas initially suggested that it may
consist of a collection of molecular 
``bullets'' ejected by the central protostar
and traveling along the outflow axis \citep{bac90}. 
Free-traveling bullets with the observed 
internal velocity dispersion of the EHV gas,
however,
would likely dissipate quickly along their path
\citep{ric92}, and because of this, 
a number of alternative interpretations
to the EHV gas have been offered.
One possibility is that the EHV gas represents shocked
ambient material accelerated by an invisible jet,
as suggested by observations of SiO emission \citep{bac91}
and by numerical simulations of momentum transfer
from a jet
\citep{che94}. Alternatively, the EHV features could represent
internal shocks along a jet 
owing to variations in the velocity of ejection,
similar to those invoked to explain the chains of HH
objects often seen in optical jets 
\citep{rag90,rag93,dut97,rei01}.

Recent high angular resolution observations of
very young outflows have started to 
offer new clues to the nature of the EHV gas and
its relation to the outflow shells.
An ideal target for this work has been the outflow from
IRAS 04166+2706 (I04166 hereafter), 
a   0.4~L$_\odot$  YSO in Taurus that presents a
number of similarities with the L1448 outflow,
including a shell-plus-jet geometry and 
bright EHV features \citep{taf04,san09}. 
High angular resolution observations of the
I04166 outflow by \citet{san09} show that the
shells and the EHV component have very different 
kinematics.
The low-velocity shells seem to arise  from
ambient gas accelerated by a wide-angle wind, 
while the
EHV gas presents a 
sawtooth velocity field that is inconsistent with 
an origin in shocked ambient material, but is
well explained as resulting from a series of
internal working surfaces traveling 
along a collimated jet. The I04166 data, therefore, suggests
that the outflow consists of two distinct components
having different physical origin.

A number of recent theoretical models have also suggested
that outflows consist of multiple components
\citep{sha06,ban06,mac08,tom10}. 
Of particular interest for the study of the EHV gas
is the so-called ``unified model'' of \citet{sha06}.
These authors have modeled the
interaction between the wind from a protostar
and the surrounding dense gas, and
suggested that while the outflow shells originate from
ambient gas that has been shock-compressed by a wide-angle wind,
the jet-like EHV component represents the central and most
collimated part of the protostellar wind itself.

Further testing the possible double-nature of the outflow gas
can be done through chemical analysis.
The chemical composition of a gas
is a sensitive indicator of its thermal
history, so if
the outflow shell and jet-like components have different physical
origins, this should be reflected in the components having 
different chemical abundances.
The chemical composition of the outflow gas has been studied
in a number of systems, both towards 
high mass protostars like
Orion-IRc 2 (e.g., \citealt{bla87})
and towards low mass objects like NGC1333-IRAS 4, L1157, or BHR 71
\citep{bla95,bac97,gar98}.
These studies have shown that the outflow composition
is characterized by abundance enhancements of species
like SiO and CH$_3$OH, which are indicative of shock
chemistry \citep{van98}. 
Previous work, however,
has not included the
very young outflows that have EHV gas, and as a result,
the chemistry of this outflow component has not
been surveyed in detail and 
compared with the composition of the rest of the outflow.
To fill this observational gap, and to test the possible 
double nature of the outflow material, we
have carried out a systematic molecular
survey of the outflow gas towards the two 
objects with most prominent EHV components, L1448 
and I04166. 

\section{Target selection and observations}

\begin{figure}
\centering
\resizebox{7cm}{!}{\includegraphics{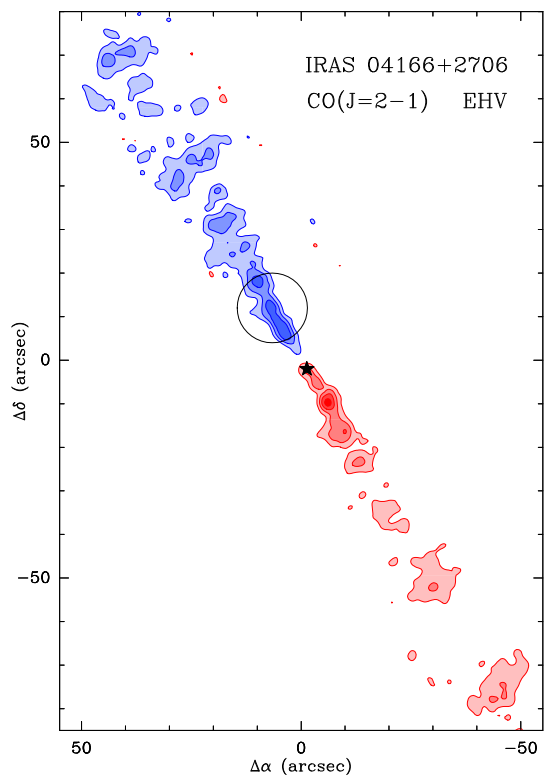}}
\caption {CO(J=2--1) map of the extremely high velocity (EHV) 
component in the I04166 outflow
illustrating the location of the molecular survey position
with a circle of $16''$ diameter (the
typical size of the telescope beam width). The CO map is 
from the
Plateau de Bure observations presented in \citet{san09}.
\label{fig_i04166_point}}
\end{figure}

The intensity of the EHV feature in high density
tracers rarely exceeds 0.1~K and is often significantly weaker,
so a molecular survey of the EHV gas requires 
integration times often in excess of
1 hour per transition. This strongly limits
the number of targets that can be studied during
a typical observing run, and for this
reason, our molecular survey was targeted towards only
two outflow positions, one in L1448 and the other in I04166.
These positions were
selected for having the brightest
CO(2--1) EHV component in each outflow, and were determined 
during a preliminary search along each outflow axis.
For the L1448 outflow, we selected the position with
offsets ($16''$, $-34''$) with respect to the driving source  
($\alpha(J2000)=3^h25^m38\fs9,$ $\delta(J2000)=+30^\circ44'05''$;
\citealt{cur90}),
which corresponds to the redshifted lobe of the flow.
For the I04166 outflow, we selected the position with
offsets ($8''$, $14''$) with respect to the mm-continuum peak 
($\alpha(J2000)=4^h19^m42\fs5,$ $\delta(J2000)=+27^\circ13'36''$;
\citealt{san09}), corresponding to the
blue lobe of the outflow. To illustrate its location,
we present in Fig.~\ref{fig_i04166_point} the I04166
target position superposed to
a high resolution CO(2--1) map of the EHV gas

In addition to the two survey targets, supplementary
positions of the L1448 and I04144 outflows were observed for this project. 
After detecting bright SO emission in the EHV components of 
the L1448 and I04166 outflows, we searched for this molecule
towards a number of  positions along the axis of each flow, to 
asses whether our target positions were unique
in presenting bright SO and to determine
the extent of the SO emission in the EHV gas.
Also, in order to derive beam-dilution factors
to correct the observed intensities during the multi-line
analysis of  Section 3, we
made small CO Nyquist-sampled maps around the two target
positions.

\begin{table}
\caption[]{Main transitions observed$^{(1)}$.
\label{tbl-lines}}
\centering
\begin{tabular}{l c c c c}
\hline
\noalign{\smallskip}
\mbox{Transition} & \mbox{Frequency}  &
\mbox{A$_{ul}$}  & \mbox{E$_{u}$/$k$} & \mbox{g$_{u}$}
\\
& \mbox{(GHz)} & 
\mbox{(s$^{-1}$)} & \mbox{(K)} &
\\
\noalign{\smallskip}
\hline
\noalign{\smallskip}
 \mbox{CO(J=1--0)} &     115.3  & $7.20 \; 10^{-8}$ &  5.54 &  3 \\
 \mbox{CO(J=2--1)} &     230.5 &  $6.91 \; 10^{-7}$ &   16.6 &  5 \\
 \mbox{$^{13}$CO(J=2--1)} &  220.4 &  $6.08 \; 10^{-7}$ &    15.9 &  5 \\
 \mbox{SiO(J=2--1)}  &      86.8 &  $2.93 \; 10^{-5}$ &    6.26 &  5 \\
 \mbox{SiO(J=3--2)}  &     130.3 &  $1.06 \; 10^{-4}$ &   12.5  &  7 \\
 \mbox{SiO(J=5--4)}  &     217.1 &  $5.20 \; 10^{-4}$ &   31.3  &  11 \\
 \mbox{SiO(J=6--5)}  &     260.5 &  $9.12 \; 10^{-4}$ &   43.8  & 13 \\
 \mbox{SO(J$_{\mathrm{N}}$=3$_2$--2$_1$)} & 99.3 &  $1.13 \; 10^{-5}$ &    9.23 & 7 \\
 \mbox{SO(J$_{\mathrm{N}}$=2$_3$--1$_2$)} & 109.3 &  $1.08 \; 10^{-5}$ &   21.1 &  5 \\
 \mbox{SO(J$_{\mathrm{N}}$=4$_3$--3$_2$)} & 138.2 &  $3.17 \; 10^{-5}$ &   15.9  & 9 \\
 \mbox{SO(J$_{\mathrm{N}}$=6$_5$--5$_4$)} & 219.9 &  $1.34 \; 10^{-4}$ &   35.0  & 13 \\
 \mbox{CH$_3$OH(J$_k$=2$_{-1}$--1$_{-1}$)E} & 96.7 & $2.56 \; 10^{-6}$ &  4.65
&  5 \\
 \mbox{CH$_3$OH(J$_k$=2$_{0}$--1$_{0}$)A} &  96.7 & $3.41 \; 10^{-6}$  & 6.97 &
  5 \\
 \mbox{CH$_3$OH(J$_k$=2$_{0}$--1$_{0}$)E} &  96.7 & $3.41 \; 10^{-6}$  & 12.2 &
  5 \\
 \mbox{CH$_3$OH(J$_k$=3$_{0}$--2$_{0}$)E} & 145.1  & $1.23 \; 10^{-5}$  & 19.2 &
  7 \\
 \mbox{CH$_3$OH(J$_k$=3$_{-1}$--2$_{-1}$)E} & 145.1 & $1.10 \; 10^{-5}$ & 11.6 &
 7 \\
 \mbox{CH$_3$OH(J$_k$=3$_{0}$--2$_{0}$)A} & 145.1  & $1.23 \; 10^{-5}$  & 13.9 &
  7 \\
 \mbox{CH$_3$OH(J$_k$=5$_{-1}$--4$_{-1}$)E} & 241.8 & $5.81 \; 10^{-5}$ & 32.5 &
  11 \\
  \mbox{CH$_3$OH(J$_k$=5$_{0}$--4$_{0}$)A} & 241.8  & $6.05 \; 10^{-5}$  & 34.8
&  11 \\
\mbox{HC$_3$N(J=10--9)} &   91.0 &   $5.81 \; 10^{-5}$ &   24.0 &   21 \\
\mbox{HC$_3$N(J=11--10)}  &   100.1 &    $7.77 \; 10^{-5}$ &    28.8 &   23 \\
\mbox{HC$_3$N(J=12--11)} &    109.2  &   $1.01 \; 10^{-4}$ &   34.1 &   25 \\
\mbox{HC$_3$N(J=16--15)} &   145.6 &  $2.42 \; 10^{-4}$ &  59.4 &   33 \\
 \mbox{CS(J=2--1)}   &      98.0 &  $1.68 \; 10^{-5}$ &    7.06 &  5 \\
 \mbox{CS(J=3--2)}   &      146.7 &  $6.07 \; 10^{-5}$ &    14.1 &  7 \\
 \mbox{CS(J=5--4)}   &     244.9 &  $2.98 \; 10^{-4}$ &   35.3  & 11 \\
 \mbox{HCO$^+$(J=1--0)} &      89.2 &  $4.19 \; 10^{-5}$ &    4.28 &  3 \\
 \mbox{HCN(J=1--0)}  &      88.6 &  $2.41 \; 10^{-5}$ &    4.26 &  3 \\
 \mbox{H$_2$CO(J$_{K_aK_c}$=2$_{02}$--1$_{01}$)} & 145.6 & $7.81 \; 10^{-5}$ 
& 10.5   &  5 \\
 \mbox{SO$_2$(J$_{K_aK_c}$=3$_{13}$--2$_{02}$)} & 104.0 & $1.01 \; 10^{-5}$ & 7.75 & 7 \\
 \mbox{H$_2$S(J$_{K_aK_c}$=1$_{10}$--1$_{01}$)} & 168.8 & $2.68 \; 10^{-5}$ & 27.9 & 9 \\
 \mbox{SiS(J=5--4)}     &   90.8 &  $1.19 \; 10^{-5}$ &    13.1 & 11 \\
 \mbox{HNC(J=1--0)}     &   90.7 &  $2.69 \; 10^{-5}$ &    4.35 &  3 \\
\hline
\end{tabular}
\begin{list}{}{}
\item[] (1) Molecular parameters from the
Cologne Database for Molecular Spectroscopy
({\tt http://www.astro.uni-koeln.de/cdms/}),
see \citet{mue01,mue05}
\end{list}
\end{table}

\begin{figure*}[t]
\centering
\resizebox{17cm}{!}{\includegraphics{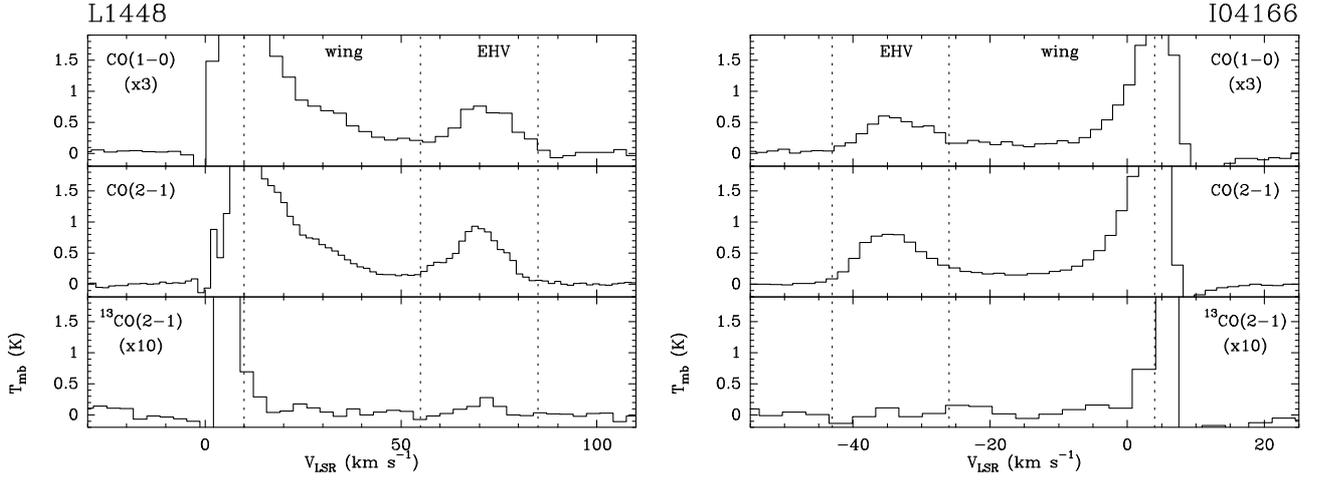}}
\caption {CO(J=1--0), CO(2--1), and $^{13}$CO(2--1) spectra towards
the target positions of the molecular survey
in the L1448 (left) and I04166 (right) outflows. The intensity scale
of the spectra
has been truncated to better show the two 
velocity components of the outflow,
the ``wing'' and the EHV regime, whose limits are indicated by
vertical dotted lines.
Negative features near the ambient velocity
of each source
($V_{\mathrm{LSR}}=5.0$~km s$^{-1}$ in L1448 and
6.7~km s$^{-1}$ in I04166)
are artifacts of the wobbler switch observing mode,
and do not affect the outflow emission studied here.
\label{fig_ehv_co}}
\end{figure*}

All observations presented here
were carried out with the IRAM 30m radio telescope
near Granada (Spain) during sessions in 2007 June, 
2008 August, and 2009 May-June. A number of
receiver configurations was used 
to maximize the number of molecular lines observed.
The list of the main transitions, rest frequencies, and related parameters 
is given in Table~\ref{tbl-lines}.
For each line, two spectrometers were used in parallel:
the VESPA autocorrelator, 
configured to provide velocity resolutions of around 1 km s$^{-1}$,
and either the 1 MHz filterbank or the wider (4 MHz) WILMA autocorrelator
to provide extra frequency coverage.
Flat baselines were achieved using
wobbler-switching mode, chopping with a frequency of 0.5 Hz 
between the source position and a reference position  
$220''$ away in azimuth.
The relatively small offset between source and reference position
caused some contamination in the spectra at the ambient cloud velocity,
but it did not affect the outflow emission studied here.
As a result of the flat baselines, only a zeroth-order polynomial
was subtracted from the spectra in the off-line data reduction.

During the observations, the pointing was checked and corrected 
by making cross scans on nearby continuum sources
approximately every two hours, and it was usually below $3''$ rms. 
A number of larger pointing excursions occurred during the 2008 run, and 
observations with suspicious pointing errors were discarded.
Calibration was achieved by observing every 10 minutes
a sequence of blank sky, an absorber at
room temperature, and a cold load, and the calibrated 
$T_\mathrm{A}^*$ scale from the telescope 
was converted into the $T_\mathrm{mb}$ scale using the
standard beam efficiency factors provided by
the telescope system.
Further reduction and analysis of the data was carried out 
using the
GILDAS software ({\tt http://www.iram.fr/IRAMFR/GILDAS}).

\section{Molecular survey results}

\subsection{CO data: wing and EHV regimes}

Figure~\ref{fig_ehv_co} shows the CO(1--0), CO(2--1) and $^{13}$CO(2--1) 
spectra towards the two target positions of our 
survey. As can be seen, each CO spectrum 
presents a broad wing component
that starts at the ambient cloud velocity
($V_{\mathrm{LSR}}=5.0$~km s$^{-1}$ in L1448 and
6.7~km s$^{-1}$ in I04166) and extends towards the red in the
case of L1448 and towards the blue in I04166.
This wing component decreases gradually in intensity 
without any sign of discontinuity until it merges
at high velocities with the bright EHV component.
This EHV component appears in the spectrum 
as a separate secondary peak of intensity
0.5-1~K and centered around $V_{\mathrm{LSR}}=70$~km~s$^{-1}$ 
in L1448 and -35~km~s$^{-1}$ in I04166. 
In L1448, the EHV component
is detected just above the noise level in the $^{13}$CO(2--1)
spectrum, while in I04166, the EHV feature can only be seen in the
main isotopologues.

To carry out our abundance analysis, we need to
define velocity limits for the wing and EHV components of each outflow,
and we use for this the CO spectra in Fig.~\ref{fig_ehv_co}.
For the EHV regime, we define its limits from the velocities at which 
this feature dominates the emission in the spectrum:
$V_{\mathrm{LSR}}$ from 55 to 85~km~s$^{-1}$
in L1448 and -26 to -43~km~s$^{-1}$ in I04166
(vertical dotted lines in Fig.~\ref{fig_ehv_co}).
For the wing regime, we take as high velocity limit 
the lower
value of the EHV component, as there is no clear gap
between the two emissions in the spectra. 
Selecting the low velocity limit for 
the wing regime requires an additional 
consideration, as the wing emission merges smoothly with the
optically thick ambient component.
To guarantee that our wing contribution is optically thin in CO, 
we have calculated the ratio between
the CO(2--1) and $^{13}$CO(2--1) spectra
and determined
the velocity at which the ratio becomes higher than 50.
This ratio is approximately equal to the expected $^{12}$C/$^{13}$C
abundance ratio in the local ISM ($\approx 57$, \citealt{lan90}),
and is comparable
to the maximum value measured with our data 
due to the limited S/N ratio of the $^{13}$CO(2--1)
spectrum.
According to our estimate, it
is reached at $V_{\mathrm{LSR}} = 10$~km~s$^{-1}$
in L1448 and at 4~km~s$^{-1}$ in I04166, which from now
on will be used as lower velocity limits for the (optically
thin) wing regime.

Although the above velocity limits help separate the
outflow into its two main components,
velocity is not the only property
that defines the wing or the EHV regime.
Interferometer maps of the CO emission 
show that each regime corresponds to a different part of the 
outflow, with the wing emission arising from a shell and
the EHV component lying along a jet (e.g. \citealt{san09}).
These two components
may therefore appear to overlap in velocity towards certain 
directions, but this does not necessarily imply that they 
spatially coexist in the outflow. Indeed, we will see below 
that the wing component seems to continue to higher velocities
inside the EHV regime.
Such an extension is very weak, and
can be usually ignored when estimating the EHV contribution. It can however
be detected in species with weak EHV emission, like HCN and possibly CS,
and in those cases, it needs to be handled with some care.

\subsection{Overview of the molecular survey}

\begin{figure*}[t]
\centering
\resizebox{17cm}{!}{\includegraphics{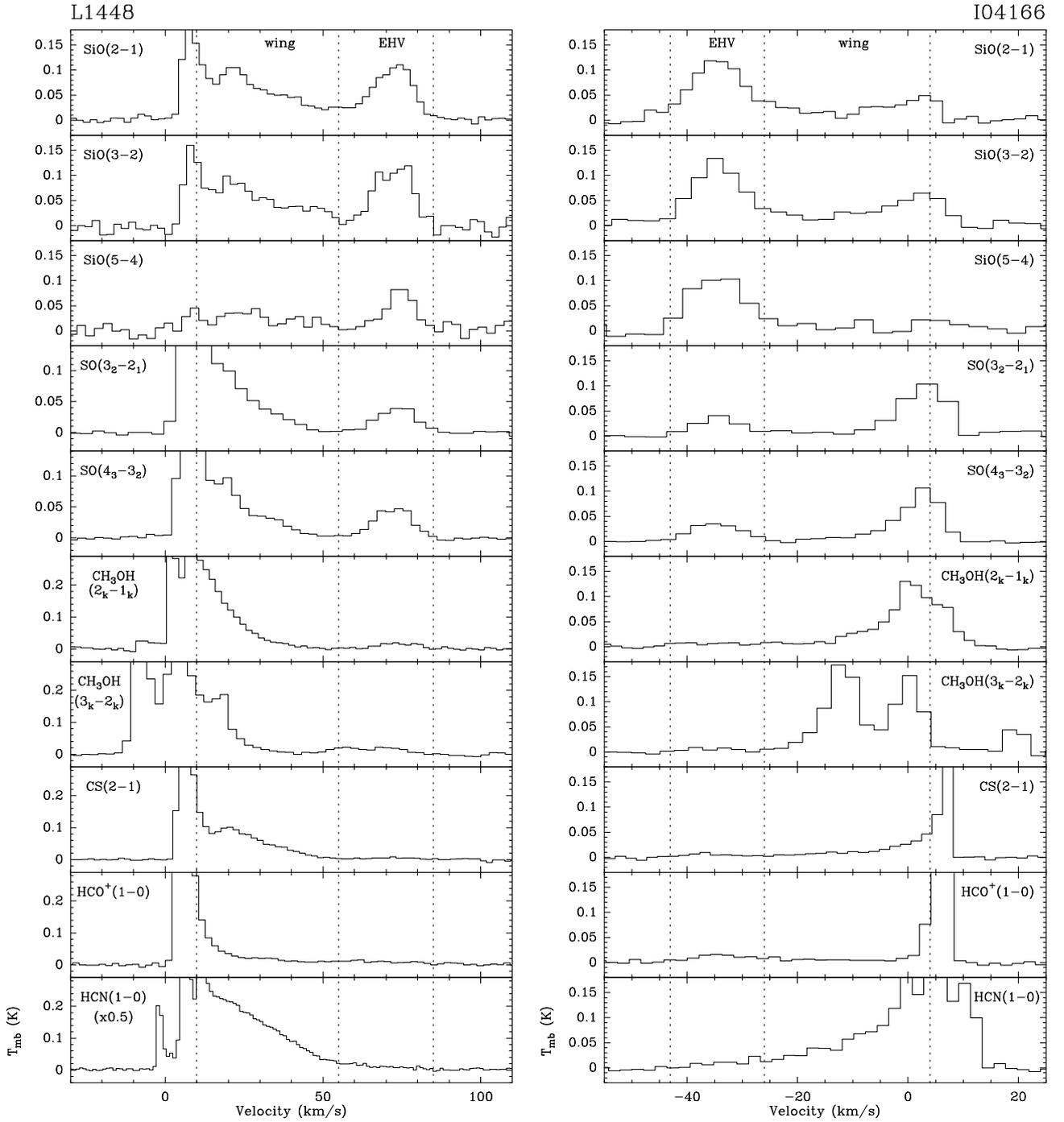}}
\caption {Spectra from the high density tracers with detected outflow
emission in the L1448 (left) and I04166 (right) survey targets. 
As in Fig.~\ref{fig_ehv_co}, the vertical dotted
lines mark the boundaries of the wing and EHV outflow regimes.
Note the prominent EHV features in the SiO and SO lines, and 
the prominent wings of lines from other species like HCN.
Note also how 
the outflow regimes partly overlap in the
multicomponent spectra of CH$_3$OH. The 
analysis used to disentangle these components is 
described in Appendix C.
\label{fig_survey}}
\end{figure*}

Fig~\ref{fig_survey} shows the spectra of the different high
density tracers detected 
towards the L1448 and I04166 outflow target positions.
As can be seen, a number of molecules
have detectable emission in one or both outflow regimes, and
overall, the spectra of most species resemble 
the CO spectra discussed in the previous section.
Bright emission in the wing regime is seen, among others, in the lines of
HCN, SO, and in the low J transitions of SiO, while weaker but still
noticeable 
wing emission can be seen in the spectra of CH$_3$OH, CS, and HCO$^+$.

As Fig~\ref{fig_survey} also shows, several species present bright
emission in the EHV regime. The SiO lines display the brightest 
EHV features, and have peak intensities of $\sim 100$~mK.
EHV SiO emission has previously been detected both in L1448 and I04166, and
this molecule has been so far the only known tracer of the
EHV gas apart from CO \citep{bac91,nis07,san09}.
As our observations indicate, SiO is not the only dense gas
tracer detectable in this outflow component.
The SO molecule, for example, also presents significant EHV emission 
in both L1448 and I04166. EHV components were detected
in SO(J$_{\mathrm{N}}$=3$_2$--2$_1$) and 
SO(J$_{\mathrm{N}}$=4$_3$--3$_2$)
with an intensity of $\approx 50$~mK, or about half of that
of SiO, and in the
brighter L1448 outflow, the higher excitation 
SO(J$_{\mathrm{N}}$=6$_5$--5$_4$)
transition was also detected at a $4 \sigma$ level (not shown).
Additional species with definitive detection of EHV emission
are CH$_3$OH (L1448 and likely I04166) and H$_2$CO (L1448 only).
They present intensities on the order of 10 mK, and their discussion
is deferred to section 3.4, where a number of weak detections
and upper limits for the EHV gas will be further studied.

Before starting our analysis of
the outflow chemical composition, we note that 
two main results from our survey can already be inferred 
from the direct inspection of the spectra in Fig.~\ref{fig_survey}.
The first result is that despite their factor of
10 difference in energetics, the L1448 and I04166 outflows
present very similar excitation and 
chemical composition in both the wing 
and the EHV components. This can be noticed 
from the similar trends seen in Fig.~\ref{fig_survey} 
towards the two outflows.
For example, in SiO, both outflows present an
EHV emission with
all transitions up to J=5--4 having similar intensities
($\approx 100$~mK) while, again in both outflows,
the wing emission fades quickly in the high-J lines. 
Also, both outflows present a
combination of relatively weak CS and HCO$^+$ wings and 
very prominent HCN wing. This all suggests that the pattern
of abundances in the two objects is very similar.

A second result that can be derived from an inspection of the 
line profiles 
is that the wing and the EHV components differ significantly in their
chemical composition. This can be inferred from a lack of correlation
between the intensity of the wing emission and the detection or not 
of an EHV
feature. An extreme example of this behavior is the HCN(1--0) line, which 
has stronger wing emission than transitions of similar Einstein-A
coefficient, like those of SiO and SO, but that has no detectable 
EHV secondary feature in the spectrum
(the HCN emission in this range seems only the continuation of
the wing, see section 3.4). Such a different behavior of the
two outflow components is a first indication that they
have undergone different chemical processing, which is
a hint of a possible different physical origin of
the wing and the EHV gas. Before
further discussion, however, we need to investigate 
the internal chemical structure of the wing component.

\subsection{Two chemical regimes in the outflow wing component}

\begin{figure*}[t]
\centering
\resizebox{17cm}{!}{\includegraphics{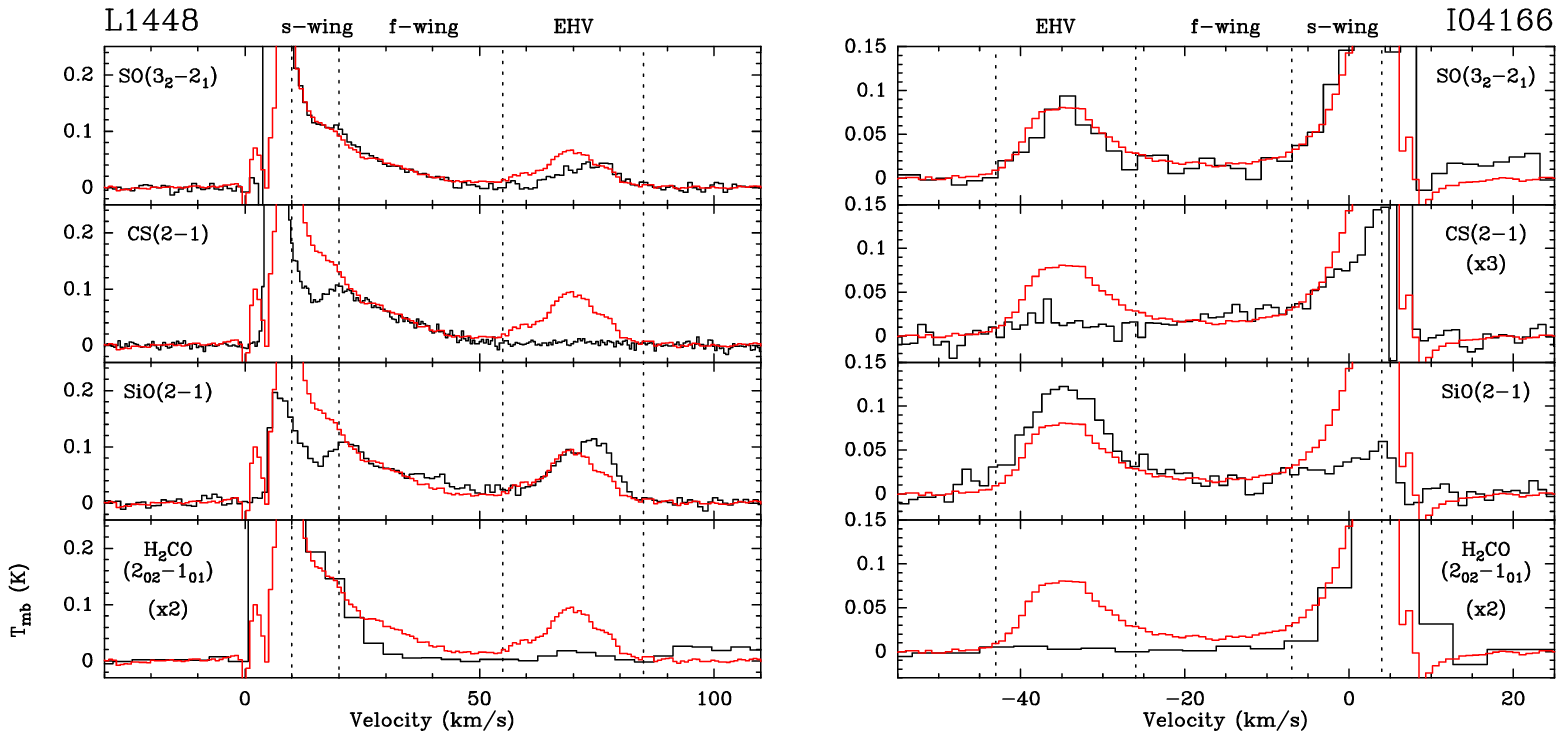}}
\caption {Comparison of spectra from selected dense gas tracers 
(black) with scaled-down versions of the CO(2--1) spectrum
(red) to illustrate the existence of two different 
chemical components in the 
wing regime of the L1448 and I04166 outflows. Note how in the
SO, CS, and SiO spectra, the fast wing (f-wing) component matches
well the CO spectrum, while the  
slow wing (s-wing) component follows a different trend: 
in SO, it matches the CO spectrum, while in 
CS and SiO, it lies below it.
In contrast, the H$_2$CO wing matches the CO
s-wing but lies below the f-wing at higher velocities.
This different behavior of the spectra in the s-wing and f-wing 
regimes suggests that the composition of these two parts of
the outflow follow different trends and need to be studied
separately
\label{fig_wing}}
\end{figure*}

The outflow wing appears in the CO spectra as a single component 
with no internal structure, but in
species like SiO, CS, or H$_2$CO, the wing presents a more complex shape.
To illustrate this behavior, we present in Fig.~\ref{fig_wing}
spectra for a few representative species
together with scaled-down versions of the CO line profile.
As can be seen, especially in the higher S/N L1448 data, 
the lines present different wing shapes, 
and only the SO spectrum resembles the CO profile over the whole
wing regime. The CS and SiO profiles, 
present two different
wing sections: a low velocity part in which the intensity
lies below the (scaled-down) CO spectrum and a high
velocity part in which the intensity follows the CO line
shape (a vertical dashed line has been drawn to separate these
two wing sections). In contrast with SO and SiO, H$_2$CO
presents a different (and unique) behavior. Its
line profile
seems to match the CO spectrum in the low velocity section
of the wing while it drops below CO at higher velocities.
As Fig.~\ref{fig_wing} shows, despite the variety of 
line profiles and wing sections, the behavior of 
each species is remarkably similar in both outflows.
This suggests that the line changes we see in the figure 
are more related to the peculiar chemistry of each species
than to the details of the outflow.

Another remarkable property of the line profiles
in Fig.~\ref{fig_wing} is the transition
between the two wing sections. For a given outflow, 
this transition occurs at the same velocity for
all species, and often in a relatively sudden way. 
In L1448, both CS and SiO spectra have a 
$\approx 5$~km s$^{-1}$-wide
velocity section just 
below $V_{\mathrm{LSR}}$ = 20~km~s$^{-1}$
where the wing shape reverses, and the
intensity of the emission increases slightly with velocity
before reaching the faster wing section
where it matches the CO profile.
It is unlikely that this dip in the wing emission arises from 
self-absorption by ambient gas because the dip is red-shifted by
10~km~s$^{-1}$ with respect to the ambient velocity. The dip is also unlikely
to arise from contamination by the reference position because it 
was seen under different observing conditions and in different
epochs. It must represent a real feature of the emission.
In I04166, no sharp transition is seen between the two wing sections, but
the CS and SiO spectra seem to match
the fast part of the CO wing at the same
velocity near $V_{\mathrm{LSR}}$ = -7~km~s$^{-1}$.

\begin{table}
\caption{Velocity ranges of the slow wing (s-wing), fast wing (f-wing), and 
extremely high velocity (EHV) regimes.
\label{tbl-vels}}
\centering
\begin{tabular}{l c c c}
\hline
\noalign{\smallskip}
\mbox{Target position} & \mbox{s-wing}  & \mbox{f-wing} &
\mbox{EHV} \\
& \mbox{(km s$^{-1}$)} & \mbox{(km s$^{-1}$)} & \mbox{(km s$^{-1}$)} \\
\noalign{\smallskip}
\hline
\noalign{\smallskip}
\mbox{L1448 ($16''$, $-34''$)$^{(1)}$} & [10, 20] & [20, 55] & [55, 85] \\
\mbox{I04166 ($8''$, $14''$)$^{(2)}$} & [-7, 4] & [-26, -7] & [-43, -26]  \\
\hline
\end{tabular}
\begin{list}{}{}
\item[] (1) Offsets with respect to $\alpha(J2000)=3^h25^m38\fs9,$
$\delta(J2000)=+30^\circ44'05''$; (2) Offsets with respect
$\alpha(J2000)=4^h19^m42\fs5,$ $\delta(J2000)=+27^\circ13'36''$.
\end{list}
\end{table}

The above differences in the shape of the profiles
indicate that for most molecules, the 
ratio of intensity with respect to CO
changes with velocity over the wing regime.
Barring excitation and optical depth effects
(discussed in the following sections), such a change
in the intensity ratio
must correspond to a change in the ratio of the abundance 
with respect to CO.
The wing regime, therefore, seems not made of
gas having uniform composition, but to consist of at
least two regimes with different chemical properties.
To include this behavior in our  abundance analysis,
from now on 
we will consider the two wing regimes separately,
and we will refer to them
as the slow wing  or {\bf s-wing} and the fast wing 
or {\bf f-wing}. Their
velocity limits, and those of the EHV 
component, are indicated by vertical dotted lines
in Fig.~\ref{fig_wing} and summarized 
in Table~\ref{tbl-vels}.

\subsection{Molecular emission in the EHV regime}

\begin{figure*}[t]
\centering
\resizebox{17cm}{!}{\includegraphics{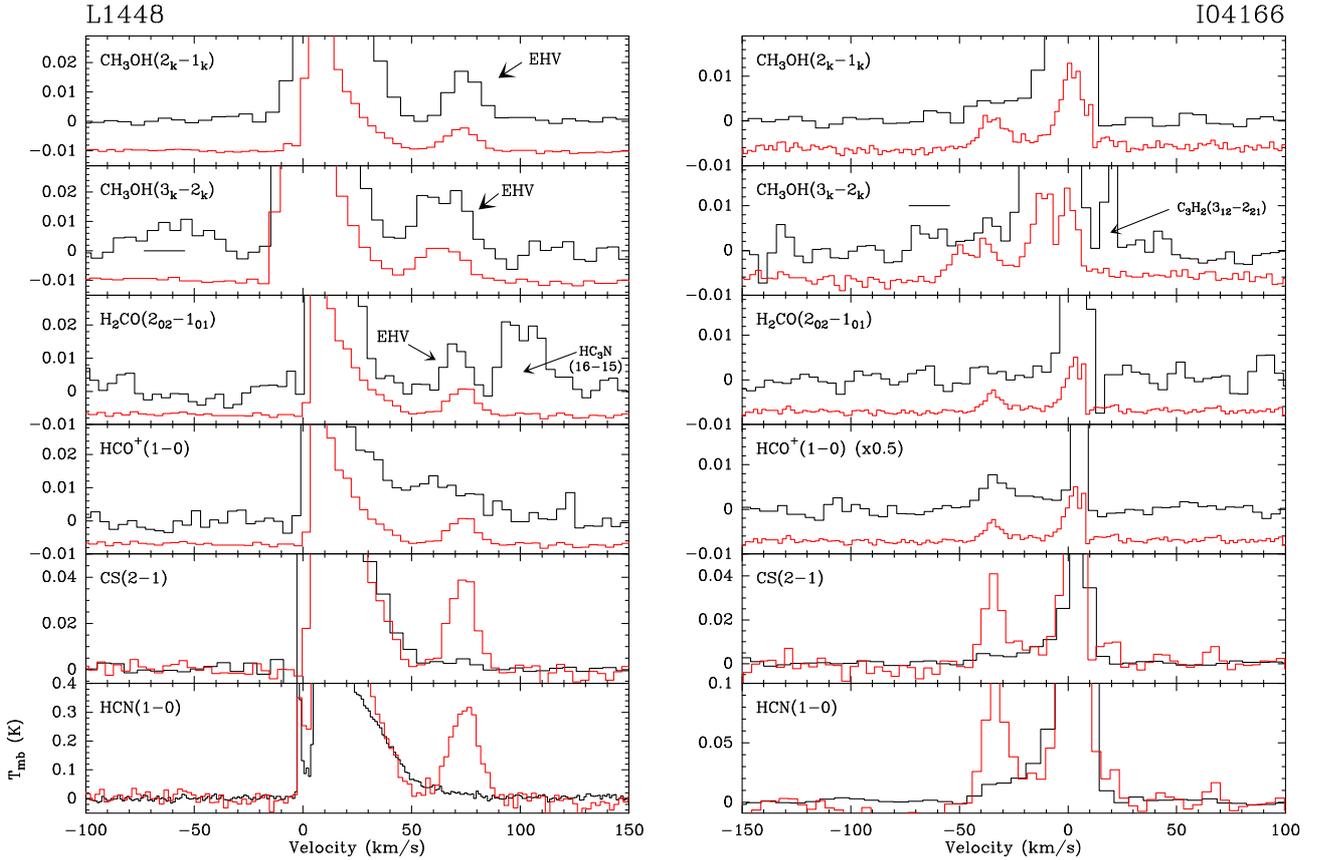}}
\caption {Evidence for EHV emission in CH$_3$OH and H$_2$CO
towards the L1448 outflow, together with possible 
detections in HCO$^+$, CS, and HCN. The black histograms
represent the data and the red histograms are
scaled-down SO(3$_2$--2$_1$) spectra
to illustrate the expected location of the EHV peak 
(near  $V_{\mathrm{LSR}}=70$~km~s$^{-1}$
in L1448 and near -35~km~s$^{-1}$ in I04166).
For CH$_3$OH and HCN, the red histogram is a 
synthetic spectrum derived using the SO(3$_2$--2$_1$)
line as a template (see Appendix C for
further details). Note the weak EHV peaks and the
gradual transition from well defined secondary
components in CH$_3$OH and H$_2$CO to more wing-like
features in HCO$^+$, CS, and HCN. 
Several unrelated lines are indicated, and 
the horizontal lines in the CH$_3$OH(3$_k$--2$_k$) 
spectra mark the position of weak $k=2$ components
that may contaminate the EHV feature in I04166.
See text for a discussion of the 
reliability of each detection.
SO was chosen as a template due to its similar behavior
to CO and higher sensitivity to dense gas.
\label{fig_ehv}}
\end{figure*}

The emission of molecules in the EHV regime deserves special attention.
Before our survey was carried out, only CO and SiO had been detected in 
this outflow component,
but as shown in section~3.2, 
SO also presents bright EHV components
in at least two transitions towards both L1448 and I04166.
The SO molecule is therefore a relatively bright 
tracer of this fastest
outflow component, and in 
Appendix A we present the results of a 
small SO survey along the axes of the L1448 and
I04166 outflows illustrating the widespread nature of the 
EHV SO emission.

In addition
to CO, SiO, and SO,
our observations revealed weak ($\sim 10$ mK) high-velocity
emission in a number of additional species, some of
them presenting a distinct EHV feature while others
of a less clear nature.
This is illustrated in Fig.~\ref{fig_ehv}, which shows
an expanded
view of the spectra for all species with
detectable emission at high velocities, together with
scaled-down versions of the SO(J$_{\mathrm{N}}$=3$_2$--2$_1$)
line to indicate
the expected position of the EHV regime. (For the 
CH$_3$OH and HCN lines, the SO models are in fact templates
of the multi component spectrum and were 
created using the method described in Appendix C.)
Given the weak emission of the features, the variety of line 
shapes, and the possible contamination from lines of
additional molecules, we will consider each line separately.

As Fig. ~\ref{fig_ehv} shows,
both the CH$_3$OH(2$_k$--1$_k$) and CH$_3$OH(3$_k$--2$_k$)
spectra towards the L1448 outflow present distinct 
secondary peaks in the velocity range 
expected for the EHV component
($V_{\mathrm{LSR}}$ = 60-80~km~s$^{-1}$.)
These peaks match closely the 
EHV features predicted by the SO template,
indicated in the figure by a 
red line plotted below the CH$_3$OH spectrum.
For CH$_3$OH(3$_k$--2$_k$), the SO template predicts
a significantly broader EHV component due to the overlap of
the contributions from the E and A CH$_3$OH species, and indeed
the observed CH$_3$OH(3$_k$--2$_k$)  EHV feature is significantly broader 
than in CH$_3$OH(2$_k$--1$_k$).
We consider this double match between observations and 
SO templates a definitive detection of CH$_3$OH 
in the EHV gas of the L1448 outflow. 

The situation in I04166 is harder to evaluate due to a significantly
(factor of 2) weaker CH$_3$OH signal and to the expected overlap 
in CH$_3$OH(3$_k$--2$_k$) between the EHV feature and a 
group of weak (mostly $k=2$) CH$_3$OH ambient-velocity components whose
position is indicated in the figure by a horizontal bar.
As Fig.~\ref{fig_ehv} shows, both CH$_3$OH spectra present significant 
emission at the expected velocity of the EHV gas, but the lack of a distinct
secondary peak isolated from the rest of the outflow wing
makes this detection less clear than that in the L1448 outflow.

Another convincing detection of an EHV feature can be seen in 
the H$_2$CO 
spectrum towards the L1448 outflow. Fig. ~\ref{fig_ehv} shows
that the H$_2$CO(2$_{02}$--1$_{01}$) line presents a 
well-defined secondary peak at the expected velocity for the EHV gas
as indicated by the SO profile, again shown in red
below the H$_2$CO spectrum.
This H$_2$CO feature  stands out  in the
spectrum as a distinct component, and is well separated from
the  brighter  peak near
$V_{\mathrm{LSR}} > 100$~km~s$^{-1}$ that
corresponds to the ambient and outflow wing
contributions from HC$_3$N(16--15).
We thus consider this feature as
a clear detection of H$_2$CO in the EHV component of the L1448 outflow.

As with CH$_3$OH, no clear H$_2$CO EHV feature can be seen
in the weaker I04166 spectrum. Our data only provides
an upper limit to the H$_2$CO emission in this outflow component, 
and a deeper integration is needed to test whether there is any
difference between the presence of H$_2$CO in the EHV component
of the two outflows.

The third species with a likely EHV feature in the spectrum 
is HCO$^+$. As Fig.~\ref{fig_ehv} shows,
the HCO$^+$(1--0) line from  the L1448 outflow presents an almost
continuous 
wing with  a hint of a relative maximum near the EHV velocity range.
In the I04166 spectrum, on the other hand, HCO$^+$(1--0) 
presents a distinct 
emission peak at the velocity expected for the EHV regime, 
suggesting that at least part of the HCO$^+$ emission 
at high velocities truly arises 
from the EHV gas. These two HCO$^+$ spectra illustrate a
trend, further seen in CS and HCN, of increasing confusion between
the EHV feature and what seems to be a high-velocity 
continuation of the wing emission. As mentioned in Sect.~3.1, the
wing emission in an outflow probably does not end at the velocity
at which the EHV component starts to dominate the spectrum, but it
continues to higher velocities overlapping with the EHV gas.
In the lines of CO (also SiO and SO), the EHV emission is brighter
than the wing, so it dominates the spectrum.
In HCO$^+$, CS, and HCN, 
the weak EHV emission barely (if at all) stands out over 
the wing emission, and its separation from the fastest
wing is ambiguous.

The last two species that we consider are 
CS and HCN. Their spectra 
present statistically significant  emission at velocities
of the EHV regime (bottom two panels in Fig.~\ref{fig_ehv}), but the nature 
of this emission is unclear. The CS(2--1) spectrum 
in both the L1448
and I04166 outflows presents rather flat emission in the
expected range of the EHV component, with a possible hint
of a secondary peak in the L1448 outflow. The I04166 spectrum 
may be slightly contaminated by a pair of C$_3$H lines, so
its shape should not be used to judge a possible EHV
peak.
The HCN(1--0) spectrum, on the other hand, presents smooth
wing-like
emission at the velocities expected for the EHV component 
(as predicted by
the SO template), suggesting that in this molecule the
wing emission overwhelms any possible contribution from
the EHV gas. Thus, while it is possible that the EHV 
component emits non-negligibly in CS(2--1), we consider
that most of the high velocity HCN(1--0) emission arises
from the extended wing component and not from the EHV gas.

In summary, our observations show {\em definite} emission
from the EHV component in CH$_3$OH and H$_2$CO towards
the L1448 outflow, {\em likely} emission 
in HCO$^+$ for both outflows, 
{\em possible} emission in CS, and {\em unlikely} emission 
in HCN. The results towards I04166 are compatible with those
towards L1448.
Given the several hours of integration needed to obtain the 
spectra of Fig.~\ref{fig_ehv}, 
it seems unlikely that a
more definitive conclusion for the marginal
cases can be reached in the foreseeable future.

\subsection{Column density estimates}

We start our abundance analysis by estimating the column density of
each species in our survey. As a first step,
we integrate the observed intensity inside each
of the three velocity ranges into which we have
divided the outflow emission (s-wing, f-wing, and EHV,
see Table~\ref{tbl-vels}).
For species like CO, SiO, SO, CS, and CH$_3$OH, we
have observed more than one transition, so we can use a
population diagram analysis to check whether the molecular energy
levels are in LTE, and if so, to derive 
an excitation temperature 
and a total column density \citep{gol99}. 
Throughout this analysis, we assume that the 
outflow emission is optically thin. For the CO 
lines the condition is guaranteed by
the comparison with the $^{13}$CO data
(section 3.1), and for
other species, the data already provides a strong hint
of the emission being thin because the observed intensities are 
on the order of 0.1~K, while the excitation temperatures are close
to 7~K (see below).

Before converting the observed line intensities into column 
densities, a number of corrections are needed.
A first correction is required to compensate for
the fact that transitions with different frequencies
were observed using different beam sizes.
As interferometer maps show (e.g., \citealt{dut97,san09}),
the outflow
emission is generally compact, especially in the EHV regime, 
so the smaller beam size of the high frequency transitions
is expected to enhance their emission. As these
lines usually correspond to high excitation levels, this
effect can introduce a 
systematic bias in the column density estimate, and
needs to be corrected. For this, 
we have used high-resolution CO data to model the dependence
of the observed intensity with beam size in each of the outflow regimes, 
and we have derived correction factors to convert
the emission of each transition into the
expected value for a $16''$ beam (the approximate mean value
of our beam sizes). The good match between line shapes of
the high density tracers and CO (Fig.~\ref{fig_wing}) suggests
that these CO-based correction factors can be applied to the 
different species in our survey.
A full description of the procedure
is presented in Appendix B.

\begin{table*}
\caption[]{Molecular column densities and excitation temperatures 
derived towards the L1448 and I04166 outflow targets}
\label{tab_lte}
\centering
\begin{tabular}{lcc|cc|cc}
\hline
\noalign{\smallskip}
& \multicolumn{6}{c}{\mbox{L1448}} \\
\hline
\noalign{\smallskip}
& \multicolumn{2}{c|}{\mbox{s-wing}}
& \multicolumn{2}{c|}{\mbox{f-wing}}
& \multicolumn{2}{c}{\mbox{EHV}}  \\
\mbox{MOLEC}  & \mbox{N$_T$ (cm$^{-2}$)} & \mbox{T$_{ex}$ (K)}
& \mbox{N$_T$ (cm$^{-2}$)} & \mbox{T$_{ex}$ (K)}
& \mbox{N$_T$ (cm$^{-2}$)} & \mbox{T$_{ex}$ (K)} \\
\noalign{\smallskip}
\hline
\noalign{\smallskip}
 \mbox{CO} & 
 $1.0 \; 10^{16} \pm 1 \; 10^{14}$ & $18.6 \pm 0.3$  & 
 $7.4 \; 10^{15} \pm 2 \; 10^{14}$ & $15.6\pm 0.5$ & 
 $5.7\; 10^{15} \pm 2 \; 10^{14}$  & $11.1 \pm 0.3$ \\
\mbox{SiO} & 
 $2.4 \; 10^{12} \pm 2 \; 10^{11}$ & $5.6 \pm 0.3$  &
 $4.7 \; 10^{12} \pm 2 \; 10^{11}$ & $7.1 \pm 0.3$ &
 $6.4\; 10^{12} \pm 2 \; 10^{11}$  & $7.0 \pm 0.3$ \\
\mbox{SO} & 
 $1.8 \; 10^{13} \pm 6 \; 10^{11}$ & $7.2 \pm 0.1$  &
 $1.4 \; 10^{13} \pm 9 \; 10^{11}$ & $7.9\pm 0.3$ &
 $1.1\; 10^{13} \pm 1 \; 10^{12}$  & $8.3 \pm 0.5$ \\
\mbox{CS} & 
 $5.5 \; 10^{12} \pm 2 \; 10^{11}$ & $6.3 \pm 0.1$  &
 $8.5 \; 10^{12} \pm 3 \; 10^{11}$ & $6.1\pm 0.2$ &
 $7\; 10^{11} \pm 3 \; 10^{11}$  & $7 \pm 1^{(1)}$ \\
\mbox{CH$_3$OH}$^{(2)}$ & 
 $7.9 \; 10^{13} \pm 1.1 \; 10^{13}$ & $10 \pm 1$  &
 $5.1 \; 10^{13} \pm 1.5 \; 10^{13}$ & $11\pm 4$ &
 $1.6 \; 10^{13} \pm 5 \; 10^{12}$  & $10 \pm 3$ \\
\mbox{HC$_3$N} & 
 $2.0 \; 10^{12} \pm 5 \; 10^{11}$ & $14 \pm 2$  &
 $2.4 \; 10^{12} \pm 1.4 \; 10^{12}$ & $11 \pm 3$ &
 $<1\; 10^{12}$  & $7 \pm 1^{(1)}$ \\
\mbox{HCO$^+$} & 
 $9.3 \; 10^{11} \pm 2 \; 10^{10}$ & $7 \pm 1^{(1)}$  &
 $7.0 \; 10^{11} \pm 3 \; 10^{10}$ & $7 \pm 1^{(1)}$ &
 $5.0\; 10^{11} \pm 4 \; 10^{10}$  & $7 \pm 1^{(1)}$ \\
\mbox{p-H$_2$CO} & 
 $1.9 \; 10^{12} \pm 3 \; 10^{11}$ & $7 \pm 1^{(1)}$  &
 $7.4 \; 10^{11} \pm 1 \; 10^{11}$ & $7 \pm 1^{(1)}$ &
 $2.3\; 10^{11} \pm 3 \; 10^{10}$  & $7 \pm 1^{(1)}$ \\
\mbox{HCN} & 
 $9.7 \; 10^{12} \pm 1 \; 10^{12}$ & $7 \pm 1^{(1)}$  &
 $1.6 \; 10^{13} \pm 3 \; 10^{11}$ & $7 \pm 1^{(1)}$ &
 $1.2\; 10^{12} \pm 2 \; 10^{11}$  & $7 \pm 1^{(1)}$ \\
\mbox{SO$_2$} & 
 $3.4 \; 10^{12} \pm 6 \; 10^{11}$ & $7 \pm 1^{(1)}$  &
 $< 2 \; 10^{12}$ & $7 \pm 1^{(1)}$  &
 $< 3\; 10^{12}$  & $7 \pm 1^{(1)}$ \\
\mbox{SiS} & 
 $< 5 \; 10^{11}$ & $7 \pm 1^{(1)}$  &
 $< 2 \; 10^{12}$  & $7 \pm 1^{(1)}$ &
 $< 1 \; 10^{12}$   & $7 \pm 1^{(1)}$ \\
\mbox{H$_2$S} & 
 $< 3 \; 10^{12}$ & $7 \pm 1^{(1)}$  &
 $< 5 \; 10^{12}$  & $7 \pm 1^{(1)}$ &
 $< 4 \; 10^{12}$   & $7 \pm 1^{(1)}$ \\
\hline
\noalign{\smallskip}
& \multicolumn{6}{c}{\mbox{I04166}} \\
\hline
\noalign{\smallskip}
& \multicolumn{2}{c|}{\mbox{s-wing}}
& \multicolumn{2}{c|}{\mbox{f-wing}}
& \multicolumn{2}{c}{\mbox{EHV}}  \\
\mbox{MOLEC}  & \mbox{N$_T$ (cm$^{-2}$)} & \mbox{T$_{ex}$ (K)}
& \mbox{N$_T$ (cm$^{-2}$)} & \mbox{T$_{ex}$ (K)}
& \mbox{N$_T$ (cm$^{-2}$)} & \mbox{T$_{ex}$ (K)} \\
\noalign{\smallskip}
\hline
\noalign{\smallskip}
 \mbox{CO} &
 $6.5 \; 10^{15} \pm 1 \; 10^{14}$ & $24.5\pm 0.6$ &
 $1.5 \; 10^{15} \pm 1 \; 10^{14}$ & $17 \pm 1$ &
 $3.2\; 10^{15} \pm 1 \; 10^{14}$ & $18\pm 1$ \\
\mbox{SiO} &
 $1.2 \; 10^{12} \pm 2 \; 10^{11}$ & $7.3 \pm 0.7$ &
 $1.1 \; 10^{12} \pm 1 \; 10^{11}$ & $8.0\pm 0.6$ &
 $4.0\; 10^{12}\pm 1 \; 10^{11}$  & $9.6 \pm 0.1$ \\
\mbox{SO} &
 $1.2 \; 10^{13} \pm 8 \; 10^{11}$ & $6.0\pm 0.2$ &
 $3.6 \; 10^{12} \pm 2 \; 10^{12}$ & $5.0 \pm 0.8$ &
 $7.3\; 10^{12} \pm 1 \; 10^{12}$  & $6.6\pm 0.4$ \\
\mbox{CS} &
 $2.1 \; 10^{12} \pm 1 \; 10^{11}$ & $5.6\pm 0.2$ &
 $1.4 \; 10^{12} \pm 3 \; 10^{11}$ & $4.6 \pm 0.5$  &
 $5.4\; 10^{11} \pm 6 \; 10^{10}$ & $7 \pm 1^{(1)}$ \\
\mbox{CH$_3$OH} $^{(2)}$ &
 $5.2 \; 10^{13} \pm 8 \; 10^{12}$  & $11 \pm 1$ &
 $8.1 \; 10^{12}  \pm 2 \; 10^{12}$  & $8\pm 2$ &
 $6.3\; 10^{12} \pm 1.8 \; 10^{12}$ & $7\pm 2$ \\
\mbox{HC$_3$N} &
 $< 2 \; 10^{12}$ & $7 \pm 1^{(1)}$ &
 $< 5 \; 10^{11}$ & $7 \pm 1^{(1)}$ &
 $< 5\; 10^{11}$ & $7 \pm 1^{(1)}$  \\
\mbox{HCO$^+$} &
 $2.6 \; 10^{11}\pm 2 \; 10^{10}$  & $7 \pm 1^{(1)}$ &
 $1.9 \; 10^{11} \pm 3 \; 10^{10}$ & $7 \pm 1^{(1)}$ &
 $3.0\; 10^{11} \pm 2 \; 10^{10}$ & $7 \pm 1^{(1)}$ \\
\mbox{p-H$_2$CO} &
 $1.8 \; 10^{12} \pm 2 \; 10^{11}$ & $7 \pm 1^{(1)}$ &
 $< 6 \; 10^{10}$ & $7 \pm 1^{(1)}$  &
 $< 5 \; 10^{11}$ & $7 \pm 1^{(1)}$ \\
\mbox{HCN}$^{(3)}$ &
 $3.2 \; 10^{12} \pm 3 \; 10^{11}$  & $7 \pm 1^{(1)}$ &
 $3.4 \; 10^{12} \pm 3 \; 10^{11}$  & $7 \pm 1^{(1)}$  &
 $2.8\; 10^{11} \pm 3 \; 10^{10}$ & $7 \pm 1^{(1)}$  \\
\mbox{SO$_2$} &
 $< 2 \; 10^{12}$ & $7 \pm 1^{(1)}$  &
 $< 1 \; 10^{12}$ & $7 \pm 1^{(1)}$  &
 $< 1\; 10^{12}$ & $7 \pm 1^{(1)}$  \\
\mbox{SiS} &
 $<5 \; 10^{11}$ & $7 \pm 1^{(1)}$  &
 $< 1 \; 10^{12}$ & $7 \pm 1^{(1)}$  &
 $< 7 \; 10^{11}$ & $7 \pm 1^{(1)}$  \\
\mbox{H$_2$S} &
 $< 5 \; 10^{12}$ & $7 \pm 1^{(1)}$  &
 $< 2 \; 10^{13}$ & $7 \pm 1^{(1)}$  &
 $< 3 \; 10^{12}$ & $7 \pm 1^{(1)}$ \\
\hline
\end{tabular}
\begin{list}{}{}
\item[Notes:] (1) $T_{ex} = 7\pm 1$~K assumed; (2) 
The E and A-forms
of CH$_3$OH were analyzed separately and the column density in the
table is the sum of the two results ($T_{ex}$ is the mean value).
The E/A ratio was found $\approx 1.2$ in 
both sources and for all outflow regimes; (3) 10\% uncertainty assumed for 
I04166 estimate of HCN column density.
\end{list}
\end{table*}
%
%

\begin{table*}
\caption[]{CO-normalized abundances $(\times 10^4)$ 
and enhancement factors with respect
to dense core abundances$^{(1)}$
\label{tab_abu}}
\centering
\begin{tabular}{lcccccc|cccccc}
\hline
\noalign{\smallskip}
& \multicolumn{6}{c|}{\mbox{L1448}} &
\multicolumn{6}{c}{\mbox{I04166}} \\

\hline
\noalign{\smallskip}
& \multicolumn{2}{c}{\mbox{s-wing}}
& \multicolumn{2}{c}{\mbox{f-wing}}
& \multicolumn{2}{c|}{\mbox{EHV}}
& \multicolumn{2}{c}{\mbox{s-wing}}
& \multicolumn{2}{c}{\mbox{f-wing}}
& \multicolumn{2}{c}{\mbox{EHV}} \\
\mbox{MOLEC}  
& \mbox{~~~X$^{\mathrm {CO}}_4$} & \mbox{f$_{enh}$}
& \mbox{~~~X$^{\mathrm {CO}}_4$} & \mbox{f$_{enh}$}
& \mbox{~~~X$^{\mathrm {CO}}_4$} & \mbox{f$_{enh}$}
& \mbox{~~~X$^{\mathrm {CO}}_4$} & \mbox{f$_{enh}$}
& \mbox{~~~X$^{\mathrm {CO}}_4$} & \mbox{f$_{enh}$}
& \mbox{~~~X$^{\mathrm {CO}}_4$} & \mbox{f$_{enh}$} \\

\noalign{\smallskip}
\hline
\noalign{\smallskip}
\mbox{SiO$^{(2)}$} & 2.4 & $4.0\; 10^3$&
6.4 & $1.1\; 10^4$&
11 & $1.9\; 10^4$ &
1.8 & $3.1\; 10^3$ &
7.3 & $1.2\; 10^4$ &
12.5 & $2.1\; 10^4$\\
\mbox{SO} & 18 & $2.7\; 10^2$  &
19 & $2.9\; 10^2$ &
19 & $2.9\; 10^2$ &
18 & $2.8\; 10^2$ &
24 & $3.7\; 10^2$ &
23 & $3.5\; 10^2$ \\
\mbox{CS} & 5.5 & 7.9 &
11 & 17 & 
1.2 & 1.7 &
3.2 & 4.7 &
9.3 & 13 &
1.7 & 2.4 \\
\mbox{CH$_3$OH} & 79 & $5.7\; 10^2$ &
69 & $5.0\; 10^2$ &
28 & $2.0\; 10^2$ &
80 & $5.8\; 10^2$ &
54 & $3.9\; 10^2$ &
20 & $1.4\; 10^2$ \\
\mbox{HC$_3$N} & 2.0 & 13 &
3.2 & 21 &
$<1.8$ & $<11$ &
$<3.1$ & $< 20$ &
$<3.3$ & $< 22$ &
$<1.7$ & $< 10$ \\
\mbox{HCO$^+$} & 0.93 & 2.0 &
0.95 & 1.9 &
0.88 & 1.8 &
0.4 & 0.8 &
1.3 & 2.5 & 
0.9 & 1.9 \\
\mbox{p-H$_2$CO$^{(3)}$} & 1.9 & 27 &
1.0 & 14 & 
0.40 & 5.5 &
2.8 & 39 & 
$<0.4$ & $<5.7$ &
$<1.4$ & $<22$ \\
\mbox{HCN} & 9.7 & 9.1 &
22 & 20 &
2.1 & 2.0 &
4.9 & 4.6 &
23 & 21 &
0.88 & 0.82 \\
\hline
\end{tabular}
\begin{list}{}{}
\item[Notes:] (1) 
X$^{\mathrm {CO}}_4 = 10^4 \times $ N$_T$/N(CO),
and $f_{enh}$ as defined in Section 4.3;
(2) The assumed SiO abundance in cores is the upper limit 
suggested by \citet{ziu89}: $5\; 10^{-12}$;
(3) An ortho-to-para ratio of 3 was assumed for H$_2$CO.
\end{list}
\end{table*}

A second correction is required to disentangle the 
overlap between multiple lines in the spectra of
HCN and CH$_3$OH.
In these spectra, the ambient lines are usually
well separated, but the outflow regimes are not.
Fortunately, the overlap between outflow components 
is only partial, and the outflow contribution to 
each individual line seems to always produce the same type of profile.
Thus, we have modeled each multi-component spectrum 
as consisting of partly overlapping lines
all having the same spectral shape 
and each centered at the velocity given by their rest frequency.
As templates for these components, we have used 
observed lines, usually from SO and SiO, and 
we have derived their relative intensities
from a fit to the full spectrum
(see Appendix C for further details).

A final consideration in the column density analysis is needed
because the
excitation temperatures derived for the outflow gas
are generally low due to subthermal
excitation ($\approx 7$~K), 
so that the common assumption 
that $T_{ex}$ is much higher than $T_{bg}$,
the cosmic background temperature, is not guaranteed.
In this situation (and assuming that the line is thin), 
the column density of the upper level of one transition ($N_u$) 
is related to the observed integrated intensity $W$ by
$$ N_u = {8 \pi k \nu^2 W \over hc^3 A_{ul}} \; 
\left[1 - {e^{h\nu/kT_{ex}}-1 \over e^{h\nu/kT_{bg}}-1}\right]^{-1},$$
where $k$ is Boltzmann's constant, $\nu$ the line frequency, 
$h$  Planck's constant, $c$ the speed of
light, and $A_{ul}$ the Einstein A-coefficient.
The term in square brackets, ignored
in the standard (high-temperature) population
diagram analysis, makes 
the column density estimate depend on $T_{ex}$, which is not known
until the analysis has been performed. Fortunately, this term is
usually close enough to 1 that $N_u$ can be solved iteratively. In
our computation, we
first assume that the term is equal to 1 and 
derive  $T_{ex}$ using the standard population analysis. Then, 
we iterate the LTE estimate recalculating $N_u$ with the $T_{ex}$ value
from the previous iteration. Convergence to a level of $10^{-4}$ in
$T_{ex}$ is usually reached in 4-5 iterations.


\begin{figure}[t]
\centering
\resizebox{\hsize}{!}{\includegraphics{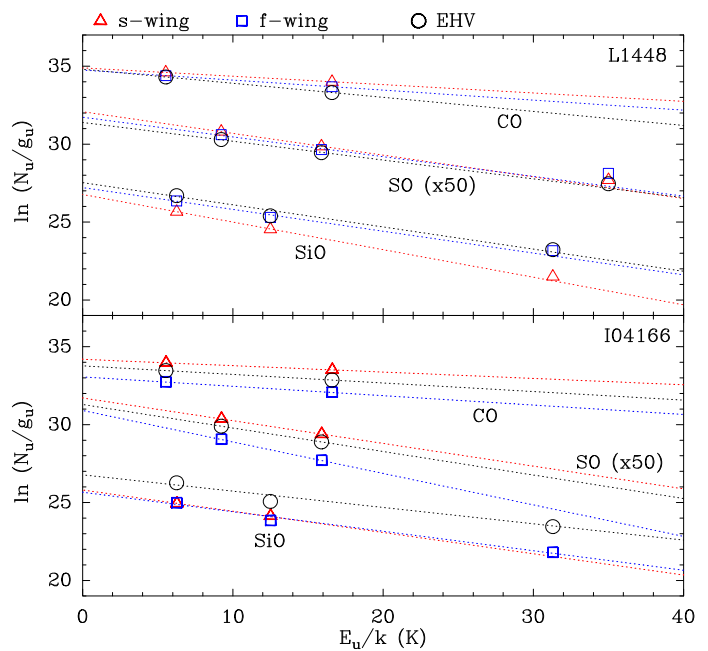}}
\caption {Population diagrams for CO, SiO, and SO in 
L1448 and I04166 
illustrating the results from the excitation analysis.
Red triangles, blue squares, and black circles 
represent s-wing, f-wing, and EHV
values, respectively. The lines show the results
from the least-squares fit described in the text
and summarized in Table~\ref{tab_lte}.
The SO data have been multiplied by 50 for presentation 
purposes only. 
\label{fig_popdiag}}
\end{figure}

As a result of our LTE analysis, we have derived 
$T_{ex}$ and column density ($N_{\mathrm{TOT}}$) values for
each outflow component in all species 
with more than one observed transition.
These values are presented Table~\ref{tab_lte}, and 
population diagrams for CO, SO, and SiO
are shown in Fig.~\ref{fig_popdiag} for illustration.
As the table and figure show, the $T_{ex}$ values
for CO are close to 15-20~K in both outflows, while the $T_{ex}$ values 
for the high-density tracers
are lower than 10~K and usually close to 7-8~K,
also in both outflows and in all the velocity regimes. 
This difference in $T_{ex}$ between CO and the other tracers 
indicates that the emission we observe in
all the outflow regimes arises from gas 
where the high-density tracers are subthermally excited
while CO is probably thermalized (if CO were
subthermal, then $T_{ex}$ for the high density tracers would
be close to the cosmic background temperature).
Such excitation conditions
are satisfied if the emitting gas has a kinetic temperature of around 
20~K and a density close to 10$^6$~cm$^{-3}$,
conditions which match closely those predicted 
by realistic models of post-shock gas
\citep{berg98}.
A more detailed analysis of the excitation conditions in
the outflow gas for a larger sample of objects will be presented in 
Santiago-Garc\'{\i}a et al. (2010, in preparation).

The similar $T_{ex}$ values shown in Table~\ref{tab_lte}
for the high-density tracers 
in the outflow suggest that for those
species for which only one transition was observed in
our survey
(e.g., HCO$^+$, H$_2$CO, HCN), we can safely assume a single
excitation value in the column density calculation. Based 
on the SiO, SO, and CS estimates, 
we take as default value $T_{ex} = 7 \pm 1$~K,
and use it to derive column densities  
for the species with only one observed transition.
For species with no detection, we use the spectrum noise level
together with the default $T_{ex}$ value to 
derive upper limits to their column density.
Table~\ref{tab_lte} summarizes all the results.

\section{Analysis of the outflow abundances}

\subsection{The similar composition of the L1448 and I04166 outflows}

\begin{figure}
\centering
\resizebox{\hsize}{!}{\includegraphics{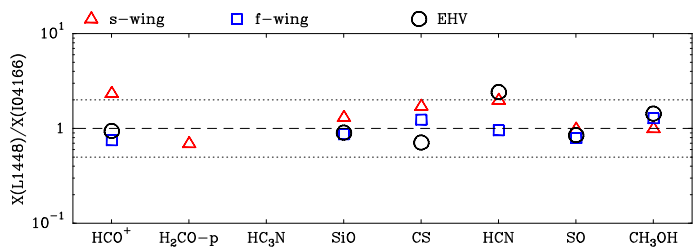}}
\caption {Ratio between the CO-normalized 
abundances derived for L1448 and I04166
in each outflow regime (symbols as in 
Fig.~\ref{fig_popdiag}). Note
how the points cluster around
a value of 1. Only detections have been used. 
No HC$_3$N data are presented due to its non 
detection in the I04166
outflow (at levels consistent with an
abundance ratio of 1, see Table~\ref{tab_lte}).
\label{fig_fig4_1}}
\end{figure}

As Table~\ref{tab_lte} shows, the CO column densities in the
L1448 outflow are
significantly higher than in I04166. The ratio between the two
is  1.5 in the s-wing regime,
4.9 in the f-wing regime, and 1.8 in the EHV regime, and these numbers
are in general agreement with the fact that the L1448 outflow is
more massive and energetic than that of I04166.
To compensate for this
difference, and to properly compare the abundance of
the different molecular species in the two
sources, we normalize each column density by the
CO value in the same regime, working from now on with
abundances with respect to CO. Such CO-normalized abundances
(referred to hereafter simply as ``abundances'' unless needed to 
avoid ambiguity)
are expected to be proportional
to the true molecular abundances (i.e., with respect to H$_2$)
given the stability of CO 
under a number of shock conditions (e.g., \citealt{berg98}).
Changes in the nature of the chemistry, however, like
from shock-processed to protostellar-wind origin, can
potentially affect the CO abundance and therefore the
normalization factor (see \citealt{gla91}).

In Table~\ref{tab_abu} we present abundance estimates with respect to
CO for all the molecules in our survey, as derived using the values
given in Table~\ref{tab_lte}. As can be seen, the 
highest CO-normalized abundances in all the
outflow regimes are those of CH$_3$OH, with 
values up to almost 10$^{-2}$ in the
wings. The next most abundant species are
SO, HCN, CS, and SiO, with values of a few times 10$^{-3}$,
also in the wing regimes. 
HC$_3$N, para-H$_2$CO, and HCO$^+$, on the other hand, present the lowest
measured abundances in our survey, with relative values respect to CO 
on the order of 10$^{-4}$. 

\begin{figure*}
\centering
\resizebox{\hsize}{!}{\includegraphics{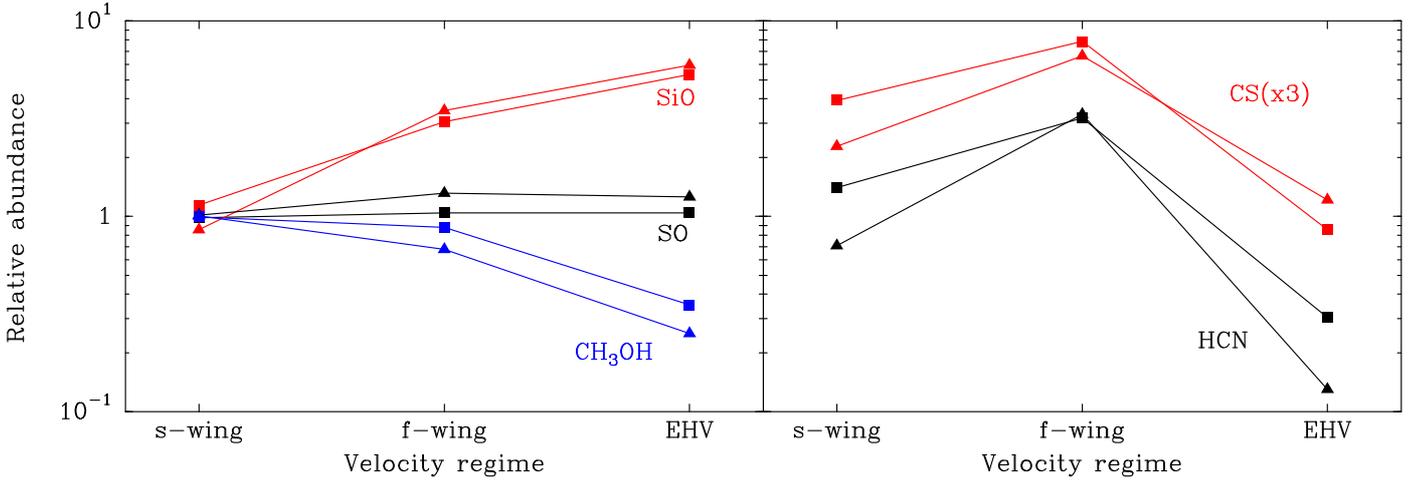}}
\caption {Variation of CO-normalized abundances 
over the three outflow regimes. Values are set to 
unity in the s-wing regime to better illustrate their change
across the outflow.
Squares represent L1448 data and triangles
represent I04166 data. Note the good agreement between the
two outflows and the systematic differences between the species.
The CS data have been multiplied by 3 to avoid overlap with the 
HCN curve. 
\label{fig_fig4_2a}}
\end{figure*}

Given the order-of-magnitude difference in energetics 
and central source luminosity 
between the L1448 and I04166 outflows, we start
our chemistry analysis by comparing the abundance results
for the two objects. This comparison is
illustrated in Fig.~\ref{fig_fig4_1} with 
the L1448/I04166 ratio of CO-normalized abundances
for each outflow regime and for
each molecular species detected in both sources.
As it can be seen, when the abundances are compared
in the same outflow regime,
all ratios are clustered around 
1 and show no clear deviation from 
that value, independent on the species or the
outflow velocity regime. The mean value of 
the ratios in the sample is 1.2 and the rms is 0.5, which suggest that
the abundances are very close, or equal, in the two sources.
This is a first indication that the abundances
determined  from  our survey may represent ``typical''
values in outflows from Class 0 sources, and not
be just peculiar to L1448 and I04166.
The comparison with the L1157 outflow abundances 
in section 4.4 will further strengthen this interpretation.

\subsection{Chemical differences between the outflow regimes}

The analysis of the spectra in Sect. 3.2
already suggested that species like CH$_3$OH and H$_2$CO  
favor low velocities, while species like SiO are enhanced
in the EHV gas. Table~\ref{tab_lte} shows quantitatively that these
trends are real, and that the abundance of most molecules
varies systematically between the 
different outflow regimes. To illustrate these 
abundance variations, 
we compare in Fig.~\ref{fig_fig4_2a} the CO-based abundances of several
molecules normalized to a value of 1 in the s-wing regime. 
As the left panel shows, the abundance of CH$_3$OH and SiO follow
opposite trends across the three outflow regimes. If the 
abundances of these species are normalized to 1 in 
the s-wing regime, they differ by a factor of 20
in the EHV gas. 
The SO abundance, on the other hand, remains approximately
constant across the three outflow regimes, indicating that
this species has either constant abundance or that 
any abundance variation follows closely that of CO.
As the figure also shows,
the abundance variations between the outflow velocity regimes 
are very similar in 
L1448 (triangles) and I04166 (squares), illustrating again the 
consistent
behavior of the chemistry in the two outflows.

\begin{figure}
\centering
\resizebox{\hsize}{!}{\includegraphics{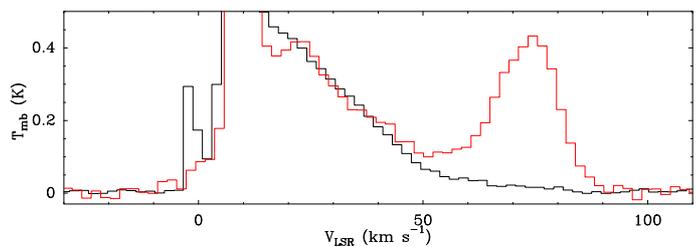}}
\caption {Comparison between the HCN(1--0) spectrum in L1448 
and a synthetic spectrum generated using SiO(2--1) as a template.
Note the good match between the two histograms 
in the f-wing and the large difference in the EHV regime. 
This change of behavior indicates a higher (factor of 20)
change in the SiO/HCN abundance ratio between
the f-wing and the EHV regime.
\label{fig_fig4_2b}}
\end{figure}

In contrast with the monotonic 
changes with velocity seen in SiO, SO, and CH$_3$OH,
the abundance of other species presents more complex 
patterns. CS and HCN, shown in the right panel of
Fig.~\ref{fig_fig4_2a}, have significant
(factor of a few) enhancements between the s-wing and
f-wing regimes, but they are strongly under-abundant 
in the EHV gas. 
As the figure shows, HCN presents the most extreme
change in the EHV regime, and its abundance drops 
by more than one order of magnitude respect to the 
f-wing value in both L1448 and I04166 
(even assuming that the wing-like 
HCN emission we observe truly represents EHV gas).
This abundance behavior of CS and HCN indicates
that the f-wing and the EHV regimes have 
rather independent chemical compositions,
and that molecules cannot be simply characterized
as favoring high or low velocities
across the whole outflow range.

Further illustration of the chemical differences
between the outflow regimes is shown in
Fig.~\ref{fig_fig4_2b}, where we compare the L1448 spectra for
HCN and SiO. These species follow a similar abundance 
increase between the s-wing and f-wing regimes, but 
they behave very differently in the EHV gas. 
To properly compare the spectra, 
we have used the SiO(2--1) line as a template to model
the HCN(1--0) hyperfine triplet, as discussed in the
Appendix C, and we have matched the intensities
of the two lines for the f-wing regime. Given the similar frequencies
of SiO(2--1) and HCN(1--0) (Table 1),
their beam-dilution factors are approximately equal,
so the different velocity regimes can be compared directly
from the spectra without additional scaling factors. As can be
seen in the figure, the spectra match well
inside the f-wing regime, but the two lines
strongly diverge from each other in the EHV gas.
SiO presents a bright EHV
peak, while HCN shows a weak wing. 
As discussed in section 3.4, the HCN wing could be a continuation of the
f-wing regime and not a separate EHV feature, so
it is difficult
to estimate the exact abundance contrast between the
two species in the EHV regime. Comparing the intensities
of the two spectra in the fastest second half of the 
EHV regime (to minimize wing contamination), 
we estimate that the SiO/HCN abundance ratio
in the EHV gas is at least a factor of 20 lower than in the
f-wing regime, and the factor could be higher if the 
high velocity wing in HCN is still part of the 
f-wing gas and does not belong to the EHV 
regime. Such a large difference
in abundance suggests that the
mechanism causing the enhancement of HCN
in the f-wing regime is not operating 
in the EHV gas, and that a combination of several chemical processes
is required to explain the abundances observed across the 
outflow regimes. Before investigating possible
processes behind the observed behavior, it is convenient to compare 
the abundances measured in the L1448 and I04166 outflows 
with those already determined for different environments.

\subsection{Comparison with dense core abundances}

\begin{figure}
\centering
\resizebox{\hsize}{!}{\includegraphics{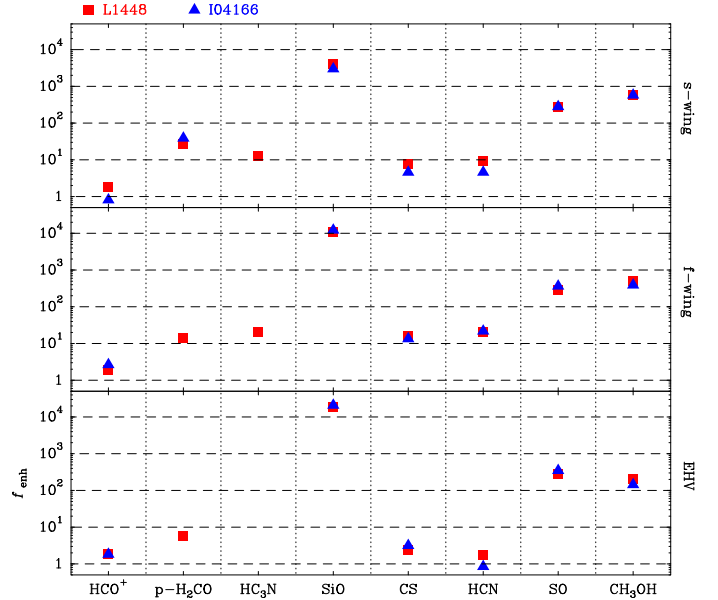}}
\caption {Abundance enhancement factors with respect to 
dense core values
for the three outflow regimes of L1448 (red squares) and I04166
(blue triangles). Note how all abundances, with the
possible exception of HCO$^+$, are significantly (order
of magnitude) enhanced in most outflow regimes. Note also
the lower enhancement of species like CS and HCN in the EHV gas.
\label{fig_fig4_3b}}
\end{figure}

We first compare the outflow abundances 
with those expected for the surrounding ambient gas.
Ideally, we should use the ambient 
line components in the outflow survey
spectra to infer the composition
of the dense gas surrounding L1448-mm and I04166.
Unfortunately, this is not possible
because the ambient components suffer
from optical depth effects and  contamination
from the reference position in the ``wobbler
switch'' observing mode.
As an alternative, we
compare the outflow abundances with 
those of similar dense cores 
for which a reliable abundance determination
exists.
Two good sources for this comparison
are the starless cores L1498 and
L1517B in the Taurus-Auriga molecular cloud (to which I04166 belongs),
whose composition has been determined in detail using multi-line
radiative transfer by \citet{taf04,taf06}.
Like most starless cores, L1498 and L1517B suffer from 
strong molecular depletion at their centers due to
freeze out onto dust grains.
For the comparison with the outflow abundances,
we use the undepleted abundances characteristic of the core
outer parts,
which most likely correspond to the pre-shock ambient
abundances for the positions in our survey.
As the abundances in L1498 and L1517B are similar,
we take their geometric mean and use the result as our reference 
value for the comparison. We define an {\em enhancement factor}
$f_{enh}$ for each species as the ratio between the outflow abundance and the
reference core abundance, and present the result for each outflow
range in Table~\ref{tab_abu}.

Fig~\ref{fig_fig4_3b} shows a plot of the $f_{enh}$ factor for each 
outflow regime in L1448 and I04166. As already mentioned, the 
L1448 and I04166 results agree with each other for all species and outflow
regimes for which we have measurements, and from now on we will
ignore source to source variations and concentrate on a few general trends
(we consider that factor-of-2 differences are at the limit of significance). 
Overall, most $f_{enh}$ factors shown in Fig~\ref{fig_fig4_3b}
are larger than unity, indicating
that the outflow abundances exceed systematically the abundances
in starless cores. As can be seen, the two
wing regimes present similar patterns of enhancement, having
the f-wing gas higher enhancement factors than s-wing
in all species but CH$_3$OH and p-H$_2$CO. The pattern of 
enhancements in the EHV gas, on the other hand, 
is clearly different from that in the wings,
especially concerning CS and HCN. As discussed in Section 4.2, 
these species are significantly
enhanced in both s-wing and f-wing regimes, 
but their abundance exceeds 
the core values only marginally in the EHV gas.

As the figure shows, the enhancement factors in the outflow
gas span 4 orders of magnitude.
HCO$^+$ is the least enhanced molecule in our survey, 
showing an overabundance of at most a factor of 2 with
respect to the dense cores, and has only marginal variations
between the different outflow regimes.
A second group of molecules consists of 
p-H$_2$CO, HC$_3$N, CS, and HCN, which present abundance enhancements
of about one order of magnitude in the two wing regimes. Interestingly,
three of these species show lower enhancement factors in the EHV gas,
while the case for HC$_3$N is unclear since its non detection
in the EHV regime imposes only a very weak constraint.
An additional order of magnitude increase in the abundance enhancement 
factor is found for SO and CH$_3$OH. As seen in the
previous section, the SO factor remains 
almost constant over the whole outflow range (at a level of
about 300), while the CH$_3$OH factor
gradually drops by about 3 from the s-wing gas to the EHV regime.
Finally, SiO presents the most extreme enhancement
factors of the sample. The SiO $f_{enh}$ factor increases 
by more than 5 between 
the s-wing and the EHV regimes, and has a typical value in the f-wing gas
of about $10^4$, with the caveat that the
non detection of SiO in starless cores makes this enhancement factor
a somewhat uncertain lower limit \citep{ziu89}. 

The above overabundance of most molecules in the outflow
gas indicates that this material has been strongly processed 
chemically with respect to the more quiescent gas around it. 
This is particularly evident in the wing component, which 
most likely arises from gas initially in the ambient 
core and that has been shock-accelerated by the outflow wind.
Indeed, the pattern of abundances in the wing components
of the outflows bears strong similarities with that 
of the L1157 outflow, as we will now see.

\subsection{Comparison with the L1157 outflow}

\begin{figure}
\centering
\resizebox{\hsize}{!}{\includegraphics{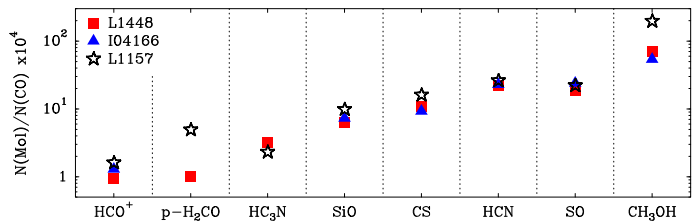}}
\caption {Comparison between CO-normalized outflow abundances
in L1448 (red squares), I04166 (blue triangles) and
L1157 (open stars) for all
species detected in the molecular survey. 
Note the good agreement (better than a factor 
of two) for most species. See text for a discussion on
the discrepancies in H$_2$CO and CH$_3$OH.
For all sources, f-wing regime values have been used.
\label{fig_fig4_3a}}
\end{figure}

The L1157 outflow is commonly considered the reference
source when studying the chemistry of low-mass outflows.
It is powered by a 11~L$_\circ$ Class 0 source \citep{ume92}, and
its composition presents high molecular abundance enhancements 
with respect to the surrounding medium suggestive of a
strong shock chemistry in action \citep{bac97}.
To compare the outflow abundances of L1448 and I04166
with those of L1157, we have re-analyzed the
data of \citet{bac97} for the so-called B1 position
(the brightest peak in the blue lobe) using the same methodology
applied in Section 3.5 to the L1448 and I04166 data. The 
results of this new analysis are not very different from those 
in the original paper, and are presented with more detail in 
Appendix D.
As the L1157 outflow lacks an EHV component, a proper comparison
with the L1448 and I04166 outflows
should be done using the wing regime.
For L1448 and I04166, we chose the
f-wing regime, since in this range
the abundance of most species remains constant with velocity,
and because for molecules with complex
spectra like HCN and CH$_3$OH, the abundances
in the f-wing regime are better determined than in the 
s-wing regime, which  is often confused by line overlaps.
For L1157-B1, we define an f-wing component taking
the fastest half of the wing emission, following 
a similar criterion to that used to define the
f-wing regime in L1448 and I04166.

Fig~\ref{fig_fig4_3a} presents a comparison between the 
CO-normalized
abundance of all the species observed both in L1157-B1
and our two outflow targets. 
As it can be seen, the abundance of most species agrees
better than a factor of 2 in the three outflows, being
the only exceptions p-H$_2$CO and CH$_3$OH, which are
more abundant in L1157. This general agreement
between the L1448 and I04166 abundances and those of L1157-B1
is remarkable given the almost 2 orders of magnitude
spanned by the values in the figure and by the 
several orders of magnitude of enhancement
with respect to the ambient values
that these abundances represent.
It is also remarkable because the L1157 outflow is very
different in physical properties from the L1448 and I04611 outflows.
In addition to lacking EHV gas, L1157 presents a 
combination of a low velocity range
and a high CO column density that makes its CO column density 
per km s$^{-1}$ higher than that of L1448 and I04166 by
factors 13 and 36, respectively.
The good matching between abundances shown in Fig~\ref{fig_fig4_3a}
therefore indicates that outflows with different physical properties,
but still powered by Class 0 sources, can have very similar chemical
compositions. It also indicates that the abundances 
do not depend strongly on
the environment where the source has formed, as
L1448 and I04166 lie in large clouds (Perseus and Taurus),
while L1157-mm was formed in a more isolated environment.
We take this good match as a suggestion that the abundances
determined in this analysis are not peculiar to the 
sources in the sample, and that they likely
represent a pattern common to other Class 0 outflows.

The similar abundances found
in  L1448, I04166, and L1157 do not
imply that there is perfect chemical uniformity.
L1157 itself presents factor-of-2 
abundance differences between its
several well studied positions (B1 versus B2, see 
\citealt{bac97,rod10}),
and high angular resolution 
observations of the B1 region reveal
internal composition gradients
\citep{ben07,cod09}.
In addition, Fig~\ref{fig_fig4_3a} shows that 
the agreement between the abundances in 
L1448, I04166, and L1157 is not perfect.
Even after ignoring abundance differences 
smaller than a factor of 2 (that could arise from
uncertainties in the analysis),  discrepancies
remain in the abundances of p-H$_2$CO and CH$_3$OH.
The factor of 5 discrepancy in the abundance of p-H$_2$CO,
however, should be treated with some care. For L1157,
only ortho H$_2$CO has been measured, while
for L1448 and I04166, only the para form was observed,
and it was necessary to assume an ortho-to-para ratio for the comparison
(the high-temperature limit of 3 was used).
Additionally, our H$_2$CO abundance determination is based
on a single lines, and it can be uncertain
due to NLTE excitation in the irregularly-spaced H$_2$CO 
energy ladder.
The factor of 3 higher CH$_3$OH abundance 
in L1157, on the other hand, seems better established, as the
same transitions were used in all sources.
Interestingly, both H$_2$CO and CH$_3$OH
are the two molecules that favor low velocities 
in L1448 and I04166 (H$_2$CO is significantly more abundant
in the s-wing regime), and they also present a similar behavior in
L1157-B1 (where CH$_3$OH is slightly more
enhanced than H$_2$CO in the slower gas).
Such velocity dependence of the abundance 
could in principle explain the higher values 
measured towards L1157, as the gas in this outflow
moves slower
than the gas in L1448 and I04166. Further observations
of  H$_2$CO and CH$_3$OH are needed to test this hypothesis,
and a more detailed study of the chemistry of these two species, 
using a  large sample of outflows, will be presented 
in Santiago-Garc\'{\i}a et al. (2010, in preparation).

Even after considering the above differences 
among outflows (and between parts of the
same outflow in the case of L1157), the main pattern 
that emerges from
Fig~\ref{fig_fig4_3a} is that of relative 
(factor of 2) homogeneity
between the composition of the different outflow wings.
As further discussed in the next section, that
composition most likely arises from shock chemistry,
and many of the observed variations are likely 
to result
from the natural dependence of this
chemistry on the shock velocity. In that 
sense, L1448, I04166, and L1157 seem 
representative of
a chemically rich early phase in the life of a
bipolar flow.

\section{Nature of the outflow regimes and their chemical composition}

\begin{figure}
\centering
\resizebox{\hsize}{!}{\includegraphics{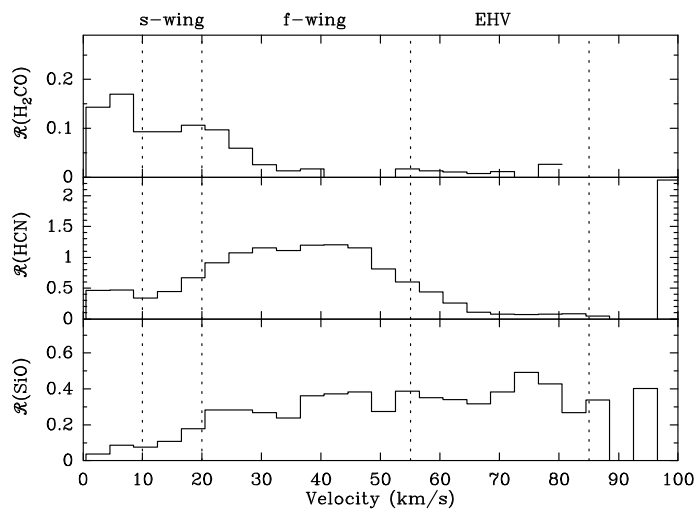}}
\caption {Intensity ratio with respect to CO in
L1448 for three species with
different behavior across the outflow regimes. 
Note how H$_2$CO favors the s-wing, HCN favors the f-wing,
and SiO is mostly present in the f-wing and EHV gas
(with a slight increase in the EHV regime).
%
\label{fig_fig4_4}}
\end{figure}

\subsection{Origin of the observed outflow velocities}

Before discussing the origin of the 
different outflow regimes, it is worth noting 
that the presence of these regimes provides by itself
useful information on the outflow velocity field.
Line spectra only provide information on the
{\em radial} component of the gas velocity field,
so they do not reveal the full 3D picture of the gas
motions. 
Outflow wings,
in particular, can arise from a variety of velocity fields,
of which there are
two extreme cases \citep{mar85}: (i) the gas has a single 
speed and the different radial velocities in the spectra
arise solely from projection effects, and (ii)  
the gas has a true distribution of speeds, and the 
observed radial velocities faithfully reflect that
distribution.
The presence of velocity-dependent 
molecular abundances can help distinguish
between these two extreme cases, as
molecular composition, in contrast to
velocity, is projection independent.
To illustrate this idea, we present in Fig.~\ref{fig_fig4_4}
the velocity-dependent emission ratio with respect to CO of 
three species, H$_2$CO, HCN, and SiO.
As can be seen, each ratio depends on velocity in a different
way: H$_2$CO is more abundant in the s-wing regime, 
HCN dominates the f-wing regime, and SiO does so over both the
f-wing and EHV gas, in agreement with the results of
previous sections.
This variety of behaviors, together with the close-to-constant 
excitation conditions derived in Sect.~3.5, 
indicates that the emitting gas
cannot consist of a single, homogeneous gas component
whose distribution in radial velocities is only due
to projection effects. The observed changes
in the gas composition as a function of velocity 
imply necessarily that the velocity regimes 
we observe are intrinsically distinct,
and that the radial velocities must represent
in some way the true (3D) velocities of the gas.

The idea that projection effects in a constant-velocity gas
cannot explain the observed outflow wings
also agrees with the high collimation
of the L1448 and I04166 outflows. Interferometer observations
of the flows
show that their opening angle near the YSO
ranges from 30$^\circ$ to 45$^\circ$, and that the
collimation at large distances is even higher than that
\citep{bac95,san09}. Projection effects from such a 
narrow range of angles seem insufficient to
explain the high velocity span of the 
outflow wings ($\sim 50$~km~s$^{-1}$ in L1448 and
$\sim 30$~km~s$^{-1}$ in I04166). Thus, in
the following discussion we will 
assume that the observed radial velocities 
are proportional to the 3D gas speed of the flow.
The proportionality factor, unfortunately, is
rather uncertain, as the inclination angle of the outflows
with respect to the line of sight is not well
constrained. For L1448, \citet{gir01} used a 2-epoch
measurement of the proper motions of the EHV SiO emission
peaks to estimate 
an inclination angle with respect to the plane of the sky
of 21$^\circ$, which implies a true EHV velocity
of about 180~km~s$^{-1}$. Although possible, the
projection correction implied by this measurement
seems very high, as
the EHV component in L1448 already has one of the highest
radial velocities observed, and with the correction, 
L1448 becomes an extreme object among its class.
Lacking an alternative measurement for I04166, and to avoid
a possible overcorrection factor, 
we will assume in the following discussion a representative 
projection angle 
of $45^\circ$ for both L1448 and I04166.
We remark, however, that a larger correction will not
invalidate the discussion below. In fact, it will make the
problems of the chemical models discussed 
to explain the EHV emission
even more acute than we estimate.

\subsection{Two outflow wing regimes: velocity-dependent shock chemistry}

As discussed before, the chemical composition of the outflow wings in
L1448 and I04166 is very similar to that 
of the L1157 outflow, which is often 
considered a textbook example of shock chemistry
\citep{bac97,ben07,cod09,rod10}.
In addition, the gas in the outflow wings is 
commonly interpreted 
as resulting from accelerated ambient gas (e.g., \citealt{san09}),
and an acceleration to the observed supersonic velocities 
necessarily implies the presence of a shock.
Thus, it seems natural to interpret 
the wing abundances in L1448 and I04166 in terms of 
shock chemistry, and to try to explain
the composition differences between the s-wing and f-wing 
gas as a result of the velocity dependence
expected for this type of chemistry.

In section 4.2 (see also Fig.~\ref{fig_fig4_4}) we saw
that the s-wing regime has lower SiO and HCN 
abundances than the f-wing gas, while it
is relatively enhanced in H$_2$CO and CH$_3$OH.
These latter molecules are often referred to as 
``primary'' species, because their abundance enhancement 
in outflows and warm gas is thought to occur via their
direct release from the mantles of dust grains
(e.g., \citealt{cha92}). In outflows, such a 
release is often explained as the result from the
sputtering of the (electrically charged) dust grains
as they stream through the neutral gas inside a C-type 
shock \citep{dra83}.
As the high H$_2$CO and CH$_3$OH abundances in the s-wing
gas indicate, grain mantle destruction must be
occurring even at the relatively low velocities characteristic
of this outflow regime, whose lower limit
we estimate in 5~km~s$^{-1}$ assuming 
a inclination angle of $45^\circ$ for both the L1448 and 
I04166 outflows.
Such low velocity starts to be uncomfortable for
grain sputtering models. The classical analysis
by \citet{dra83} predicts that most water
ice is returned to the gas phase due to sputtering for shocks
faster than 25~km~s$^{-1}$, and although the number has
been recently lowered slightly by
\citet{jim08} for the release of CH$_3$OH and H$_2$O,
it is still on the order of 20~km~s$^{-1}$.
(A much lower value, around
10~km~s$^{-1}$ was proposed for dirty ices by
\citet{flo94}, but the value of the
sputtering threshold used by these authors is uncertain.)
Requiring that our observed minimum s-wing
radial velocity
corresponds to the high speed of the
above models would imply that both outflows 
are less than $15^\circ$ from the plane of the sky, 
which seems unlikely given
the already very high (radial) velocity of their EHV gas.
On the other hand, although it is possible that the
outflow gas has slowed down significantly
from its initial shock velocity, 
Fig.~\ref{fig_fig4_4} shows that the
SiO and HCN abundances drop in the s-wing
regime compared to the f-wing regime. This suggests
that only little gas from the f-wing regime 
has been transferred to the s-wing part of the flow.

A possible solution to the observed 
low-velocity enhancement
in CH$_3$OH and H$_2$O is to invoke an additional
process of mantle disruption, like 
direct collisions between dust
grain particles \citep{tie94}. \citet{cas97}
have shown that grain-grain collisions in oblique
C-type shocks can lead to complete water ice evaporation 
at shock velocities as low as 15~km~s$^{-1}$. These authors,
in addition, suggest that the release of less bound species
like CH$_3$OH could occur at velocities as
low as 10~km~s$^{-1}$, in better agreement with our
observations.
Clearly more work is needed to understand
low-velocity mantle disruption, which is
not peculiar from the sources of our survey,
but seems occur with relatively high frequency
(e.g., \citealt{gar02}).

In contrast with the s-wing regime, the f-wing gas
is characterized by
an abundance drop of H$_2$CO (and to a less extent CH$_3$OH)
together with an enhancement of SiO, HCN, and CS (Fig.~\ref{fig_fig4_4}).
This composition change occurs at a velocity of approximately
20~km~s$^{-1}$ for both L1448 and I04166,
again assuming an outflow inclination angle of $45^\circ$.
As illustrated by the Fig.~\ref{fig_fig4_4}, the composition 
change between s-wing and f-wing regimes
is relatively sudden, especially taking into account that after 
the transition, the abundance of most species remains almost constant for 
about 20 to 30~km~s$^{-1}$ (up to the EHV regime).
Such a relatively sharp transition between the s-wing and f-wing
regimes suggests that one or more additional
chemical processes start to operate in the gas at that
velocity, and that they continue to operate for the rest of the
f-wing regime. To our knowledge, no current chemical model
can explain the changes we observe. The abundance decrease
of H$_2$CO and CH$_3$OH, for example, suggest that these
species are being destroyed at high velocities either
during grain sputtering or via gas-phase reactions, while
mantle sputtering 
and grain-grain collision models commonly predict a steady increase 
in abundance with velocity (e.g., \citealt{cas97,jim08}).
The enhancement of SiO, on the other hand, has been modeled 
in detail by
a number of authors, and two alternative scenarios
have been usually explored: release of silicon from 
grain cores and from grain mantles 
\citep{sch97,cas97,gus08a,gus08b}.
The sudden abundance increase of SiO near 20~km~s$^{-1}$ 
could in principle be interpreted as indicating the 
crossing of the threshold for core grain disruption
(although a value more 30~km~s$^{-1}$ would have
been expected, \citealt{gus08a}). Still,
the relatively constant SiO abundance in the f-wing gas 
after crossing that threshold
is puzzling, since models predict a steady increase of 
the SiO yield with velocity once grain core erosion starts.
Thus, although our observed f-wing abundances are
likely to result from shock chemistry, and the
values measured towards L1448 and I04166 
agree well with each other (and with those in L1157),
no current model seems to explain all the observed trends.

\subsection{The peculiar composition of the EHV component}

The composition of the EHV gas has so far been 
poorly known due to the lack of data 
apart from CO and SiO. 
The detection of SO, CH$_3$OH, and H$_2$CO, together with
the less clear detections or upper limits for HCO$^+$, CS, and 
HCN, make it possible now to explore the chemistry
of this gas component in more detail.
As discussed in section 4.2, the EHV
gas partly continues 
the trend of velocity-dependent abundances  
found in the wing components:
SiO becomes more enhanced,
SO (and probably HCO$^+$) remain almost constant,
and H$_2$CO and CH$_3$OH
drop slightly with respect to 
the wing gas.
CS and HCN, on the other hand, 
deviate strongly from this behavior. The two
species are significantly ($\approx$ factor of 2) enhanced
in the f-wing regime compared to the s-wing gas, but the
transition to the EHV gas represents an
abundance drop of about one order of magnitude.
The most extreme example of this behavior is HCN in L1448, due
to its high signal-to-noise ratio (Figs.~\ref{fig_fig4_2b} and 
\ref{fig_fig4_4}).
Even assuming that the weak, wing-like HCN feature
that we observe at the highest velocities 
corresponds to the EHV gas and is not the (more likely)
continuation of the f-wing regime, the 
HCN abundance in the EHV component is a factor of 10-20 
lower than in the 
f-wing gas. This drop is so high that any HCN enhancement over the
ambient abundance
observed in the wing regime seems to have disappeared in the EHV gas
(Fig~\ref{fig_fig4_3b}).

The CS and HCN abundance drop
between the f-wing and EHV regimes represents the highest contrast
seen in our survey, and suggests a significant chemical difference 
between these  two outflow regimes. HCN and CS have in common
that their enhancement in shocked gas 
depends on the presence of carbon 
in the gas phase.
HCN is mainly produced in shocks by the reaction of
N with CH$_2$, while CH$_2$ itself originates from the 
reaction of C with H$_3^+$ or C$^+$ with H$_2$
\citep{pin90,bog05}. CS, on the other hand, is 
enhanced in shocks by the reaction of SO and 
C, which like the previous reactions, has no activation
barrier \citep{pin93}. The fact that HCN and CS can be enhanced in the
wing gas but not in the EHV component 
suggests that abundance of free C in 
the EHV gas is lower than in the wing.
Such a situation could occur if 
most of the C in the EHV gas has been quickly locked up
into CO, leaving behind 
little free C for additional reactions. 
If this were so, we would expect the EHV gas to contain
a relative excess of oxygen. Indeed, 
Figs.~\ref{fig_survey} and \ref{fig_ehv}
show that all molecules with prominent
EHV feature are
O-bearing species: CO, SiO, SO, CH$_3$OH, and H$_2$CO.
(Note also that O also destroys CH$_2$, the precursor of 
HCN, \citealt{bog05}).
We take this peculiar composition of the EHV gas as evidence that this
component is richer in oxygen 
(or poorer in carbon) than the other outflow regimes.

The peculiar C/O abundance ratio in the EHV gas is a potential 
clue to the nature of this outflow component.
As discussed in the introduction, 
 we can classify models of the
EHV component as ``primary wind'' or ``shocked ambient gas,'' depending
on whether the material originates from the protostar
(or innermost vicinity) or from the surrounding ambient cloud.
Unfortunately, no detailed chemical models
exist for the composition expected in either 
interpretation of the EHV gas. 
Different aspects of the chemistry of shocks and winds,
however, have been previously studied, and we can 
use these results to explore possible differences between the
two alternative origins of the EHV component.
If the EHV gas represents ambient material originally at rest
that has been shock-accelerated by a protostellar
wind, the speed of the wind has to equal or exceed the
velocity of the EHV component.
For I04166 and L1448, the radial velocity of the 
fastest EHV gas 
with respect to the ambient cloud is
70 and 110~km~s$^{-1}$, respectively, again
assuming a standard correction for an inclination 
angle of 45 degrees.
Realistic models of molecular shocks in gas at densities
of $10^4$~cm$^{-1}$ or higher
predict that for velocities higher than 50-70~km~s$^{-1}$, the shock
will dissociate H$_2$ and will become of J (jump) type 
\citep{dra83,leb02}.
Any shocks giving rise to the EHV components of L1448 and I04166
must therefore be of J type, so the species observed in this
outflow regime must have reformed from the
dissociated material in the post-shock
gas. Although  models of J-type shocks predict 
efficient reformation of molecules
in the postshock gas, they also predict
enough UV radiation to photodissociate CO
and keep a relatively high abundance of neutral carbon.
The detailed models of J-type shocks by \citet{neu89}
show significant HCN formation in the postshock gas,
and predict HCN 
column densities that are comparable to, and often one order
of magnitude higher than those of SO
(unfortunately, these authors do not predict CS abundances).
Our observations of the EHV gas in
L1448 and I04166, however, show a reversed situation, 
where the SO column density exceeds that
of HCN by one order of magnitude or more in the
EHV component (abundances in
the f-wing gas are comparable in both outflows). 
This overproduction of HCN 
(by almost two orders of magnitude with respect to SO)
seems to pose a significant challenge to the 
interpretation of the EHV component as shock-accelerated
ambient gas. 

To explore the alternative interpretation of the EHV component 
as a protostellar wind, we use the pioneering work of
\citet{gla91}, who studied the composition of
a 150 km s$^{-1}$ wind ejected quasi spherically from the surface of
a low-mass protostar. 
(\citealt{pan09} have recently presented an alternative
model of a disk wind, but they
find no SiO production, in contrast with observations).
\citet{gla91} found that although the protostellar
wind starts atomic in composition, it can become 
molecular for high enough densities (corresponding 
to mass-loss rates of a few 10$^{-6}$~M$_\circ$~yr$^{-1}$).
Due to the complexity of the calculation, however, \citet{gla91}
only considered a limited chemical network which did not
include the chemistry of SO, HCN, or CS, 
so we can only make educated guesses on the production rates
for these species. Given the relatively high abundance 
predicted for OH in the wind, and the 
lack of activation energy for its reaction with S \citep{pin93},
SO formation in the wind seems favorable for the same
conditions for which CO and SiO are abundant. 
The possibility of HCN and CS production, on the other hand,
is harder to estimate. It is however noticeable that in the denser
models explored by \citet{gla91} 
(mass loss rates higher than $3\; 10^{-6}$~M$_\circ$~yr$^{-1}$
and cases 2 and 3 in the authors notation), all the carbon is
locked up into CO at large radii, and no C or C$^+$ is available
for further reactions (their Table 4). This lack of C suggests
that HCN and CS production may not be favorable for the same 
wind models that predict abundant CO and SiO, in agreement with
our observations of the EHV gas. (Note however that the
model overproduces SiO, but the chemistry of this species is
very sensitive to the gas density and the assumed stellar temperature.)
In addition to the unexplored chemical 
paths, the \citet{gla91} model contains 
a number of uncertainties that may affect the gas chemistry, like
the effect of the higher collimation
expected for a jet geometry and the role of the stellar UV field.
Still, its chemical paths offer a promising mechanism to explain the
observed behavior of CS and HCN in the EHV gas.

More challenging for the two scenarios of the origin of the 
EHV gas
is the presence of CH$_3$OH and H$_2$CO with abundances
significantly enhanced with respect to the
ambient gas (Sect.~4.3). 
Gas-phase production of CH$_3$OH, in particular, is highly  
inefficient \citep{gep06}, and any abundance
enhancement of this species is usually interpreted 
as resulting from the disruption of 
dust grain mantles, where CH$_3$OH had previously 
formed via gas-grain reactions \citep{wat04}.
Both shocked-ambient-material and protostellar-wind
models, however, need to rely mostly on 
gas-phase reactions to produce their molecules.
In the shocked ambient-gas model, 
molecules are formed in the post-shock
gas due to the dissociative 
nature of the J-type shock, while in 
the wind model of \citet{gla91}, the gas starts
atomic and hot because it originates 
near the protostellar surface.
Despite this reliance on gas phase reactions, it
may be possible for either model 
to produce CH$_3$OH and H$_2$CO
in the EHV gas by re-forming it on the
dust grains via successive hydrogenation of CO 
\citep{wat04}. Indeed, 
both the J-type shock and wind models
predict dust grains and 
abundant atomic hydrogen in the gas phase
\citep{neu89,gla91}. Unfortunately, no
detailed analysis of this chemical path has been
studied so far.

As the above discussion illustrates, 
much information lies still untapped in the
chemical composition of the outflow material and in
the variations with velocity that this composition
suffers. To extract this information, detailed
modeling of the chemistry in both
shock-accelerated ambient gas and primary winds
from protostars and disks
is needed. Also needed is a further observational effort
to characterize a large sample of outflows in the
species found to be sensitive to gas kinematics.
Chemical composition, due to
its sensitivity to the thermal history of the gas and 
its initial conditions, offers 
a unique window to investigate the so-far invisible processes
that take place during the production and
acceleration of a bipolar outflow.

\section{Conclusions}

We have carried out a molecular 
survey of one position in each of the outflows powered
by L1448-mm and I04166, two Class 0 YSOs
in the Perseus and Taurus molecular clouds. These two 
outflows belong to a special class whose CO spectra
presents a standard outflow wing 
and a discrete, extremely high velocity (EHV) component,
and they are believed to represent one of the earliest phases
of the outflow phenomenon.
The main results of our study are the following.

1. We have detected a number of molecules in the
wings of both outflows, including CO, SiO, SO, CS,
HC$_3$N, HCO$^+$, H$_2$CO, HCN, and CH$_3$OH.

2. In the EHV component, we 
have detected definitively SO, CH$_3$OH, 
and H$_2$CO, as the spectra from these species 
present distinct EHV secondary peaks
similar to those previously seen in CO and SiO.
In contrast, the spectra from HCO$^+$, CS, and HCN
present low-level wings that 
overlap with the EHV regime, but the
origin of this emission 
as part of a distinct EHV components is less clear.

3. For most molecules, 
we have used the intensities of multiple transitions 
together with 
a population diagram analysis to derive 
excitation temperatures and column densities.
To compensate for differences in the telescope
beam between the different transitions, we have estimated
beam-dilution factors using high angular resolution
CO maps of the outflows. 

4. When normalized to CO, the abundances in the
L1448 and I04166 outflows agree with each other 
better than a factor of 2. This suggests that the
abundance patterns we have found may be
common to other very young outflows.

5. Based on chemical composition, we identify three 
different outflow regimes.
Two regimes are part of the wing component and the third
one consists of the EHV gas. The low-velocity
wing regime presents 
velocity-dependent abundances for most species
and is richer than the others in 
CH$_3$OH and H$_2$CO. The high-velocity wing 
regime has almost-constant abundance with velocity.
and the gas is richer in species like SiO, HCN, and CS
compared to the slower wing. Finally, the
EHV gas presents a mixed composition:
it has a very high SiO abundance but is 
relatively poor in HCN and CS.

6. The outflow abundances 
typically exceed those of the Taurus dense cores by one
or two orders of magnitude. On the other hand, the fast wing
abundances are very similar to those of
L1157, the prototypical ``chemically active'' flow. 
In agreement with previous work, we interpret the
high abundances in the wing components 
as resulting from shock chemistry.
Comparing the slow and fast
wing regimes, we find
that their composition differences likely
originate from the sensitivity of
shock chemistry to the velocity of the shock,
although no existing model seems capable of 
explaining all features observed in the data.
Our observations illustrate the importance of
velocity in the final composition of 
shock-accelerated gas.

7. Concerning the EHV gas, our molecular survey 
shows its peculiar chemistry. This gas
combines high abundances of 
O-rich species together with an order
of magnitude drop in the abundance of CS and HCN 
with respect to the wing gas. This indicates
that the EHV gas has a much lower C/O
ratio than the shock-accelerated wing gas,
and we suggest that it may be a 
consequence of the two outflow components having
different physical origins. One possible
explanation of this difference is that the
EHV gas represents part of the primary wind from the protostar
or its vicinity. Further modeling of the composition
of this type of wind is needed to reach a definite
conclusion.

\appendix

\section{A survey of EHV SO emission in I04166}

\begin{figure}
\centering
\resizebox{\hsize}{!}{\includegraphics{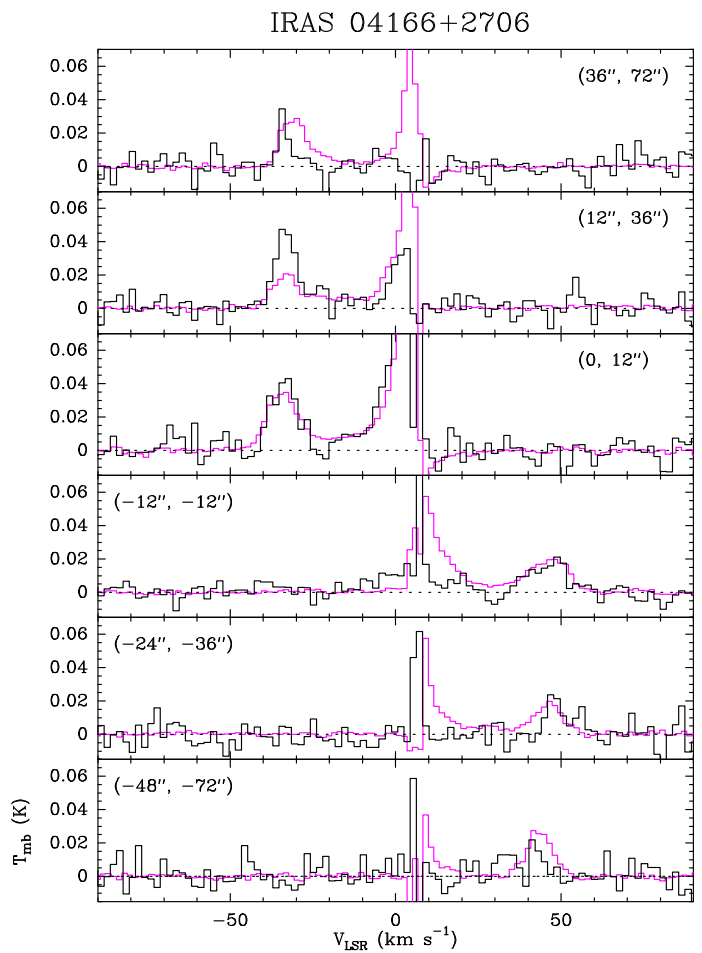}}
\caption {SO($3_2$--$2_1$) (black) and CO(2--1) (red) 
spectra along the central axis of the I04166
outflow. The CO spectra have been scaled down by a factor of 20,
and the offsets are referred to the IRAS nominal position
($\alpha(J2000)=4^h19^m42\fs6,$ $\delta(J2000)=+27^\circ13'38''$).
\label{fig_i04166_co_so}}
\end{figure}

The detection of relatively strong SO emission in the EHV component of
both L1448 and I04166
target positions was unexpected.  No SO emission
from EHV gas had previously been reported, so
no  information on its distribution and relation
to other EHV tracers like CO and SiO was available. To 
better characterize this SO emission, we 
made small maps around
both sources. In I04166, 
we observed 6 positions along the outflow axis
and in L1448 we observed three positions.
This appendix presents the I04166 data
(the more limited L1448 survey is consistent with
the I04166 results).

Fig.~\ref{fig_i04166_co_so} shows the 
SO(3$_2$--2$_1$) and CO(2--1)
spectra observed simultaneously along the
I04166 outflow axis. As can be seen, 
EHV SO emission was detected in at least
5 of the 6 target positions, both towards the red and 
blue lobes. Considering the limited S/N of the data
and the different beam size of the two transitions
($24''$ in SO(3$_2$--2$_1$) and $11''$ in CO(2--1)), 
we find a good match between the SO and
CO data, indicating that the two EHV emissions 
originate from the same gas. In this respect, we
can consider the position selected for the
molecular survey as a likely 
representative of the  outflow
as a whole. 

\begin{table}
\caption[]{EHV emission along the I04166 outflow$^{(1)}$.
\label{tbl-i04166}}
\centering
\begin{tabular}{ccc}
\hline
\noalign{\smallskip}
\mbox{Offsets$^{(2)}$} & \mbox{I[CO(2--1)]$^{(3)}$}  & 
\mbox{I[SO(3$_2$--2$_1$)]$^{(3)}$} \\
($''$,$''$) & \mbox{(K km s$^{-1})$} & \mbox{(K km s$^{-1})$} \\
\noalign{\smallskip}
\hline
\noalign{\smallskip}
 (36,72) & 5.0 & 0.12 \\
 (12,36) & 3.9 & 0.38 \\
 (0,12) & 7.7 & 0.40 \\
 (-12,-12) & 5.2 & 0.21 \\
 (-24,-36) & 3.9 & 0.13 \\
 (-48,-72) & 4.9 & 0.07 \\
\hline
\end{tabular}
\begin{list}{}{}
\item[Notes:] (1) Velocity ranges are [$-43$, $-26$] for blue EHV and
[37, 54] for red EHV; 
(2) Offsets as in Fig.~\ref{fig_i04166_co_so};
(3) typical $\sigma$(I) is 0.1 K km s$^{-1}$ for CO(2--1) and
0.03 K km s$^{-1}$ for SO(3$_2$--2$_1$).
\end{list}
\end{table}

To further characterize the EHV SO emission in the I04166
outflow, we study the ratio
between the CO(2--1) and SO(3$_2$--2$_1$) integrated intensities.
Table~\ref{tbl-i04166} presents the values measured in our survey
using the LSR velocity ranges $-43$ to $-26$ km s$^{-1}$ for the 
blue EHV gas and 37 to 54 km s$^{-1}$ for the red EHV gas.
As can be seen, the CO(2--1)/SO(3$_2$--2$_1$) ratio
in the innermost 4 positions is on the order of 20, which is consistent
with the value derived in the molecular survey of Sect. 3.
In the outermost position of each outflow lobe, however,
the SO(3$_2$--2$_1$) emission weakens significantly, 
and the CO(2--1)/SO(3$_2$--2$_1$)
ratio increases to about 40 towards the
northern (blue) lobe and twice as much towards the southern (red) lobe.

The close to constant CO/SO intensity ratio along 
the inner EHV jet suggests that
both the SO excitation and the abundance relative to CO are
close to constant for this outflow component 
over the central $\approx 0.03$~pc,
and that they have values similar to those derived for the 
($8''$, $14''$) position, for which additional transitions
of both species were observed. 
The increase of the CO/SO intensity ratio
in the outer outflow, on the other hand,
suggests that there is a drop in the SO
abundance or excitation of the EHV gas 
about 0.05~pc from the outflow source.
Lacking additional SO data, we use the CO and SiO
observations of \citet{san09} to investigate this drop.
The high-resolution CO and SiO interferometric
observations showed a similar factor of 2-3 increase in the 
CO(2--1)/SiO(2--1) ratio in the EHV component
as a function of distance from the protostar.
The interferometric data, in addition, showed
that both the CO and SiO emitting regions broaden with distance to 
the YSO, suggesting that the EHV gas is spreading laterally
as it moves away from the protostar (in agreement
with models of internal working surfaces, \citealt{dut97,san09}).
In this case, it seems therefore likely that the excitation
of SiO drops faster than that of CO due to its higher dipole
moment, and that the observed change in the CO/SiO ratio
is at least in part produced by a change in excitation.
As the SO molecule
has also a high dipole moment, it is very likely that the increase 
in the CO/SO ratio with distance from I04166 also 
results, at least in part, from a
change in excitation.
Multitransition observations 
are required to fully constrain 
the characteristics of the EHV gas in I04166. 

Although our observations seem to 
represent the first detection of SO in 
EHV gas, bright SO emission in the
jet-like components of the HH211 and Ori-S6 outflows have been 
recently reported 
by \citet{lee10} and \citet{zap10}, respectively.
The jet-like component of these outflows
has significantly lower velocity than the components in
L1448 and I04166, and their emission
does not form distinct EHV features in the
CO spectra like those shown in Fig.~\ref{fig_ehv_co}.
These jet-like 
components, however, may represent a similar phenomenon
to the EHV gas, and may owe their low radial velocity just to
a location close to the plane of the sky.
Lacking accurate measurements of outflow
inclination angles, 
a comparison between the chemical properties 
of these objects
may help test the idea that all the jet-like
components share a common physics (and likely
origin). 
If so, chemical composition may result a more
discriminant tracer of outflow properties
than gas kinematics, and 
the presence of EHV gas in outflows
may be more common among Class 0
sources than currently recognized.

\section{Correction for differential beam dilution}

\begin{figure}
\centering
\resizebox{\hsize}{!}{\includegraphics{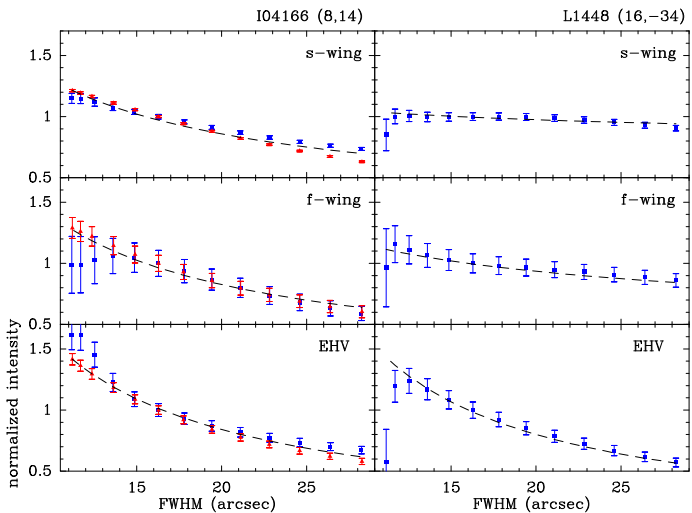}}
\caption {Variation of the CO(2--1) integrated intensity 
as a function of total convolving beam size (normalized to the $16''$ value)
for the three outflow velocity regimes used in the molecular survey.
The intensities have
been determined by convolving appropriately either IRAM 30m Nyquist-sampled
maps (blue squares) or a combined 
Plateau de Bure interferometer+30m data set (red triangles). 
The dashed lines represent a set of fits with the form
$\Omega_{\mathrm{MB}}^{-\alpha}$
that have been used to correct the survey data
for beam dilution effects (see text). 
The error bars represent the estimated 1-$\sigma$ rms and have been calculated 
from the noise in the spectra using a baseline fit to the line-free channels.
\label{effect_convolution}}
\end{figure}

The beam size of a diffraction-limited telescope
like the IRAM 30m 
depends linearly on the operating wavelength, so
for a given molecule,
transitions having different rest frequencies
are observed using different angular resolutions. 
At the CO(1--0) frequency, for example, 
the FWHM of the 30m main beam 
is about 21 arcsec, while the main beam size 
for CO(2--1) is half that value \citep{gre98}. This significant 
change in the beam size with frequency, 
together with the small angular size of
the regions under study, makes each transition 
in our survey suffer from a different amount of beam 
dilution. Such a differential effect
needs to be corrected before using the observed intensities
to derive molecular abundances,
or otherwise the high frequency (small beam) transitions
will be over weighted, producing an overestimate
of both the molecular excitation and the species column density.
In this appendix we present the method we have
used to compensate for this common observational problem.

Properly correcting all our survey data 
for the effects of beam dilution 
would require making high angular resolution maps for each 
transition. In the absence of such 
data, we used observations of the main outflow
tracer, CO, to characterize the distribution of gas in the vicinity of our 
survey targets, and to estimate how the observed spectrum 
depends on the beam size. 
To this end, we made a Nyquist-sampled map 
around each outflow target in CO(2--1), as this transition
provides an effective angular resolution higher than most 
transitions in our survey ($\approx 11''$).
This CO(2--1) map was 
used to simulate observations with larger beam sizes 
by convolving the Nyquist-sampled spectra to the required resolution.
As a result, we produced
a grid of CO(2--1) spectra with effective
beam sizes ranging from about $12''$ (close to the beam size
for CO(2--1)) to about $30''$ (larger than the largest beam  
in our survey).
To double check the I04166 results (where a pointing error
was suspected in the Nyquist map), we repeated the calculation using 
the combined Plateau de Bure-30m data set from \citet{san09},
which contains all the spatial frequencies, is independent of
our new 30m Nyquist-sampled map, and has an angular resolution
of about $3''$.

The results of our convolution analysis are presented 
in Fig.~\ref{effect_convolution} in the form of 
the CO(2--1) 
integrated intensities as a function of the 
total convolving beam size
for the three outflow regimes defined in the
molecular survey.
As can be seen, the intensity in all regimes 
drops systematically with beam size, which is
an indication that the 
emission is only partly resolved by the telescope, and
that differential beam dilution truly affects 
our survey data. As can also be seen, the
slope of the drop is different in each outflow regime,
and it steepens systematically 
from the s-wing to the EHV gas. This steepening is in good agreement with
the previous observation that outflow collimation 
increases systematically with velocity 
in both L1448 and I04166 \citep{bac90,taf04b}.
Also seen in Fig.~\ref{effect_convolution} is the
good agreement between the
two data sets of I04166 for beam sizes larger than 
about $14''.$ The slight 
disagreement for smaller beam sizes partly results 
from the high noise in the 30m spectra with small beam size, as the
Nyquist-sampled map is relatively shallow and only a few 
spectra contribute to the small beam size data points.
This small beam size regime is also very sensitive
to small pointing errors, and this seems to be the cause 
of some deviations seen in both the I04166 and L1448 
plots.

The systematic dependence of the integrated intensity with
observing beam size shown in Fig.~\ref{effect_convolution} 
suggests that the effect of beam dilution 
can be corrected, at least to first order, 
using a simple parameterization. For a symmetric Gaussian
beam, the beam dilution factor is proportional to 
$\Omega_{\mathrm{MB}}^{-2}$ 
if the source is point like, to 
$\Omega_{\mathrm{MB}}^{-1}$
if the source one dimensional, and trivially equal to 
$\Omega_{\mathrm{MB}}^{0}$
if the source is infinitely extended 
in two dimensions.
It seems therefore reasonable that when fitting the curves 
of Fig.~\ref{effect_convolution}, we use an
expression of the form 
$\Omega_{\mathrm{MB}}^{-\alpha}$,
where 
$\alpha$ is a free parameter expected to lie between 0 and 2.
This expectation is confirmed by the data, and as
the dashed curves in Fig.~\ref{effect_convolution}
show, the analytic expression fits the data well
inside the range of observations.
The $\alpha$  coefficient in the fit is 0.6, 0.75, and 0.9
for the s-wing, f-wing, and EHV ranges of I04166, and
0.1, 0.3, and 1.0 for the same ranges in L1448, respectively.
Of particular interest is the close-to-1 values derived for
the EHV range in both outflows, a further indication
that this fastest outflow regime is highly collimated.

With the above 
$\Omega_{\mathrm{MB}}^{-\alpha}$
fits we have 
converted all measured integrated intensities into equivalent
values for a $16''$ beam, which is the average size 
of the beams used in 
in our survey. The excitation and column density estimates 
presented in this paper therefore represent estimates of
these parameters inside a $16''$-diameter beam.

\section{Fitting multiple-component spectra with templates}

The spectra of HCN and CH$_3$OH contain multiple
components that partly overlap, and to calculate
the integrated intensity of each of these components
in each outflow regime,
we need a method to disentangle them.
An inspection of the data shows that in each 
multi-component spectrum,
all components present similar shape and differ only in
velocity and relative intensity, and this suggests that
it should be possible to disentangle the outflow 
contributions 
by fitting the complex spectrum with a synthetic one 
made of multiple copies of the same line profile.
To carry out such a fitting procedure, we have
taken spectra from single lines (from CO, SO, CS, or SiO)
as templates,
made copies of them using the CLASS software, and
shifted each template in frequency by the appropriate value
for each component
(as provided by the Cologne Database for Molecular Spectroscopy, 
see \citealt{mue01,mue05}).
By varying the relative intensity of each template, which is
the only free parameter in the procedure, we have fitted 
the multi-component spectra, and from this fit, 
we estimate the integrated
intensity of each line component in the three outflow
regimes (only the outflow
part was fitted while the often optically thick ambient
emission was ignored).

To identify the best fit solution, we selected a set of
velocity ranges in the spectrum where the overlap 
between components is minor
or where one component clearly dominates the emission,
and we required that the synthetic spectrum matched the
observed integrated intensity within the noise level.
For the case of HCN(1--0), we found that the
three hyperfine components had relative intensities very
close to their statistical weights, suggesting that the 
outflow emission is optically thin
and that the excitation of the sublevels is close to LTE. 
By fixing the relative intensities of the components 
to their statistical
weights, the only one free parameter left to fit the 
observed emission 
profile was a global scaling factor.

As shown in Fig.~\ref{fig_wing}, different species present
slightly different outflow components, reflecting their 
peculiar abundance pattern across the outflow regimes (Section 4.2).
For this reason, some species provide better
templates than others to fit a given multi-component spectrum, 
and finding the best match is necessary a matter of trial and error. 
To test how sensitively the final integrated intensities 
depend on the choice of template species, we fitted the 
HCN(1--0) line from L1448 (the multiple-component spectrum 
with highest S/N in our sample) using templates from
CS(2--1), SO(23--12), CO(2--1), and SiO(2--1).
When these templates are properly scaled to match
the observed spectrum, the rms
dispersion in the estimate of the 
outflow contribution to the
s-wing, f-wing, and EHV regimes
is 12\%, 2\%, and 14\%, respectively. 
If these dispersions are estimates of the
uncertainty introduced by the multi-component fitting
procedure, we conclude that the added uncertainty is
comparable to
the uncertainty in the telescope calibration and in the correction
for beam dilution. The template method, therefore,
seems to provide an accurate way to
estimate the outflow contribution in multiple-component
profiles.

\section{Revised abundances for L1157-B1}

To properly compare the abundances of our target
outflows with those of the L1157-B1 position, often considered
as a standard in outflow chemistry, 
we have re-evaluated the L1157-B1 estimates of
\citet{bac97} for the species detected in both L1448 and I04166.
We have used the original data from \citet{bac97} and we
have applied to it the same procedure used in
the analysis of L1448 and I04166:
correction for beam dilution, use of templates to disentangle
overlapping components, and use of 
population diagram analysis for species
with more than one detected transition.
In contrast with \citet{bac97}, who integrated the intensities
across the full line profile, we have ignored
LSR velocities less than 1.8~km~s$^{-1}$ away from
the systemic velocity ($V_{\mathrm{LSR}} = 
2.3$~km~s$^{-1}$) to avoid possible contamination
from the ambient cloud. This velocity limit 
was set by the point at which the $^{13}$CO(2--1)
emission drops to the noise level in the spectrum.
As in L1448 and I04166, we have
divided the line wing into two
regimes that will be referred to as the s-wing and the 
f-wing (see limits in Table~\ref{tab_l1157}).
For each outflow regime, we have derived beam-dependent dilution factors
as described in Appendix B,
using the CO(2--1) data from \citet{bac97}, and we have 
fitted the results with power laws 
having $\alpha$ values of 0.6 and 0.7 for s-wing and f-wing, 
respectively.

The results of the abundance re-analysis
are presented in Table~\ref{tab_l1157}. To better
determine the relatively high $T_{\mathrm ex}$ value 
of CO ($\approx 50$~K),
we have complemented the J=1--0 and 2--1 data of \citet{bac97}
with the 3--2, 4--3, and 6--5 data from \citet{hir01},
also corrected for beam-dilution effects.
For species with only one observed transition, we have used
a default $T_{\mathrm ex}$ value of 12~K as determined from
the population diagram analysis of SiO, CS, and SO.
Overall, the results of our re-analysis are in good
(factor of 2) agreement with those of \citet{bac97}
when the abundances are normalized to that of CO. A
comparison between the values for the s-wing and f-wing regimes
shows that, as in L1448 and I04166, SiO
becomes relatively more abundant at high velocities,
while CH$_3$OH and H$_2$CO decrease in abundance with velocity.
The other species re-analyzed here present similar
abundances within 20 \% in the two outflow regimes.

\begin{table}
\caption[]{Molecular column densities and excitation temperatures
for  L1157-B1
\label{tab_l1157}}
\centering
\begin{tabular}{lcc|cc}
\hline
\noalign{\smallskip}
& \multicolumn{2}{c|}{\mbox{s-wing$^{(1)}$}}
& \multicolumn{2}{c}{\mbox{f-wing$^{(1)}$}} \\
\mbox{MOLEC}  & \mbox{N$_T$} & \mbox{T$_{ex}$}
& \mbox{N$_T$} & \mbox{T$_{ex}$} \\
& \mbox{(cm$^{-2}$)} & \mbox{(K)}
& \mbox{(cm$^{-2}$)} & \mbox{(K)} \\
\noalign{\smallskip}
\hline
\noalign{\smallskip}
 \mbox{CO} & $3.2 \; 10^{16}$ & 51 & $1.1 \; 10^{16}$ & 52 \\
\mbox{SiO} & $1.9 \; 10^{13}$ & 12  & $1.1 \; 10^{13}$ & 12 \\
\mbox{SO} & $8.1 \; 10^{13}$ & 17  & $2.3 \; 10^{13}$ & 15 \\
\mbox{CS} & $4.8 \; 10^{13}$ & 10  & $1.7 \; 10^{13}$ & 12 \\
\mbox{CH$_3$OH$^{(2)}$~~~~~} & $1.0 \; 10^{15}$ & 11  & $2.1 \; 10^{14}$ & 11 \\
\mbox{HC$_3$N} & $8.2 \; 10^{12}$ & 26  & $2.5 \; 10^{12}$ & 23 \\
\mbox{HCO$^+$} & $5.3 \; 10^{12}$ & $12^{(3)}$  & $1.7 \; 10^{12}$ & $12^{(3)}$ \\
\mbox{o-H$_2$CO} & $6.4 \; 10^{13}$ & $12^{(3)}$  & $1.6 \; 10^{13}$ & 
$12^{(3)}$ \\
\mbox{HCN} & $9.8 \; 10^{13}$ & $12^{(3)}$  & $2.9 \; 10^{13}$ & $12^{(3)}$ \\
\hline
\end{tabular}
\begin{list}{}{}
\item[Notes:] (1) $V_{\mathrm{LSR}}$ limits are 
$-3.5$ to 0.5~km~s$^{-1}$ for s-wing and $-7.5$ to $-3.5$~km~s$^{-1}$
for f-wing; (2) The E and A-forms
of CH$_3$OH were analyzed separately and the column density in the
table is the sum of the two results. The E/A ratio was found 
to be $\approx 1.2$ in both outflow regimes; (3) $T_{ex} = 12$~K was 
assumed.
\end{list}
\end{table}

\begin{acknowledgements}
We thank the staff of the IRAM 30m telescope for help
during the observations, and Paola Caselli and Eric Herbst
for enlightening conversations on shocks and chemistry.
This research has made use of NASA's 
Astrophysics Data System Bibliographic Services and the SIMBAD 
database, operated at CDS, Strasbourg, France. 
This work was partially supported by MICINN, within the program
CONSOLIDER INGENIO 2010, under grant ``Molecular Astrophysics:
The Herschel and ALMA era - ASTROMOL'' (ref.: CSD2009-00038).
\end{acknowledgements}


\begin{thebibliography}{}
\bibitem[Bachiller et al.(1990)]{bac90} Bachiller, R., Cernicharo, J.,
Mart\'{\i}n-Pintado, J., Tafalla, M., \& Lazareff, B.\ 1990, \aap, 231, 174 
\bibitem[Bachiller et al.(1991)]{bac91} Bachiller, R., 
Mart\'{\i}n-Pintado, J., \& Fuente, A.\ 1991, \aap, 243, L21 
\bibitem[Bachiller et al.(1995)]{bac95} Bachiller, R., Guilloteau, S.,
Dutrey, A., Planesas, P., \& Mart\'{\i}n-Pintado, J.\ 1995, \aap, 299, 857
\bibitem[Bachiller \& P\'erez Guti\'errez(1997)]{bac97} Bachiller, R., 
\& P\'erez Guti\'errez, M.\ 1997, \apjl, 487, L93 
\bibitem[Bachiller \& Tafalla(1999)]{bac99} Bachiller, R., \& Tafalla, 
M.\ 1999, NATO ASIC Proc.~540: The Origin of Stars and Planetary Systems, 227 
\bibitem[Banerjee \& Pudritz(2006)]{ban06} Banerjee, R., \& Pudritz, R.~E.\ 
2006, \apj, 641, 949
\bibitem[Bally \& Stark(1983)]{bal83} Bally, J., \& Stark, A.~A.\ 1983, 
\apjl, 266, L61 
\bibitem[Benedettini et al.(2007)]{ben07} Benedettini, M., 
Viti, S., Codella, C., Bachiller, R., Gueth, F., Beltr{\'a}n, M.~T., 
Dutrey, A., \& Guilloteau, S.\ 2007, \mnras, 381, 1127 
\bibitem[Bergin et al.(1998)]{berg98} Bergin, E.~A., Neufeld, 
D.~A., \& Melnick, G.~J.\ 1998, \apj, 499, 777 
\bibitem[Blake et al.(1995)]{bla95} Blake, G.~A., Sandell, 
G., van Dishoeck, E.~F., Groesbeck, T.~D., Mundy, L.~G., 
\bibitem[Blake et al.(1987)]{bla87} Blake, G.~A., Sutton, 
E.~C., Masson, C.~R., \& Phillips, T.~G.\ 1987, \apj, 315, 621 
\& Aspin, C.\ 1995, \apj, 441, 689 
\bibitem[Boger \& Sternberg(2005)]{bog05} Boger, G.~I., 
\& Sternberg, A.\ 2005, \apj, 632, 302 
\bibitem[Caselli et al.(1997)]{cas97} Caselli, P., Hartquist, T.~W., 
\& Havnes, O.\ 1997, \aap, 322, 296 
\bibitem[Charnley et al.(1992)]{cha92} Charnley, S.~B., 
Tielens, A.~G.~G.~M., \& Millar, T.~J.\ 1992, \apjl, 399, L71 
\bibitem[Chernin et al.(1994)]{che94} Chernin, L., Masson, 
C., Gouveia dal Pino, E.~M., \& Benz, W.\ 1994, \apj, 426, 204 
\bibitem[Codella et al.(2009)]{cod09} Codella, C., et al.\ 2009, 
\aap, 507, L25 
\bibitem[Curiel et al.(1990)]{cur90} Curiel, S., Raymond, 
J.~C., Moran, J.~M., Rodr\'{\i}guez, L.~F., \& Cant\'o, J.\ 1990, 
\apjl, 365, L85 
\bibitem[Draine \& McKee(1993)]{dra93} Draine, B.~T., \& McKee, C.~F.\ 
1993, \araa, 31, 373 
\bibitem[Draine et al.(1983)]{dra83} Draine, B.~T., Roberge, 
W.~G., \& Dalgarno, A.\ 1983, \apj, 264, 485 
\bibitem[Dutrey et al.(1997)]{dut97} Dutrey, A., Guilloteau, S., \& Bachiller, 
R.\ 1997, \aap, 325, 758 
\bibitem[Flower \& Pineau des Forets(1994)]{flo94} Flower, D.~R., 
\& Pineau des Forets, G.\ 1994, \mnras, 268, 724 
\bibitem[Garay et al.(1998)]{gar98} Garay, G., 
K{\"o}hnenkamp, I., Bourke, T.~L., Rodr{\'{\i}}guez, L.~F., 
\& Lehtinen, K.~K.\ 1998, \apj, 509, 768 
\bibitem[Garay et al.(2002)]{gar02} Garay, G., Mardones, D., 
Rodr{\'{\i}}guez, L.~F., Caselli, P., 
\& Bourke, T.~L.\ 2002, \apj, 567, 980 
\bibitem[Geppert et al.(2006)]{gep06} Geppert, W.~D., et al.\ 
2006, Faraday Discussions, 133, 177
\bibitem[Girart \& Acord(2001)]{gir01} Girart, J.~M., \& Acord, J.~M.~P.\ 
2001, \apjl, 552, L63 
\bibitem[Glassgold et al.(1991)]{gla91} Glassgold, A.~E., 
Mamon, G.~A., \& Huggins, P.~J.\ 1991, \apj, 373, 254 
\bibitem[Goldsmith \& Langer(1999)]{gol99} Goldsmith, P.~F., \& Langer, W.~D.\ 1999, 
\apj, 517, 209 
\bibitem[Greve et al.(1998)]{gre98} Greve, A., Kramer, C., \& Wild, W.\ 1998, 
\aaps, 133, 271 
\bibitem[Gusdorf et al.(2008a)]{gus08a} Gusdorf, A., Cabrit, S., 
Flower, D.~R., \& Pineau Des For{\^e}ts, G.\ 2008, \aap, 482, 809 
\bibitem[Gusdorf et al.(2008b)]{gus08b} Gusdorf, A., Pineau Des For{\^e}ts, 
G., Cabrit, S., \& Flower, D.~R.\ 2008, \aap, 490, 695 
\bibitem[Hirano et al.(2010)]{hir10} Hirano, N., Ho, 
P.~P.~T., Liu, S.-Y., Shang, H., Lee, C.-F., 
\& Bourke, T.~L.\ 2010, arXiv:1005.0703 
\bibitem[Hirano \& Taniguchi(2001)]{hir01} Hirano, N., \& Taniguchi, Y.\ 
2001, \apjl, 550, L219 
\bibitem[Hollenbach \& McKee(1980)]{hol80} Hollenbach, D., \& McKee, 
C.~F.\ 1980, \apjl, 241, L47 
\bibitem[Jim{\'e}nez-Serra et al.(2008)]{jim08} Jim{\'e}nez-Serra, I., 
Caselli, P., Mart{\'{\i}}n-Pintado, J., \& Hartquist, T.~W.\ 2008, 
\aap, 482, 549 
\bibitem[J{\o}rgensen et al.(2007)]{joe07} J{\o}rgensen, 
J.~K., et al.\ 2007, \apj, 659, 479 
\bibitem[Langer \& Penzias(1990)]{lan90} Langer, W.~D., \& Penzias, A.~A.\ 
1990, \apj, 357, 477 
\bibitem[Le Bourlot et al.(2002)]{leb02} Le Bourlot, J., 
Pineau des For{\^e}ts, G., Flower, D.~R., 
\& Cabrit, S.\ 2002, \mnras, 332, 985 
\bibitem[Lee et al.(2010)]{lee10} Lee, C.-F., Hasegawa, 
T.~I., Hirano, N., Palau, A., Shang, H., Ho, P.~T.~P., 
\& Zhang, Q.\ 2010, \apj, 713, 731 
\bibitem[Machida et al.(2008)]{mac08} Machida, M.~N.,
Inutsuka, S.-I., \& Matsumoto, T.\ 2008, \apj, 676, 1088
\bibitem[Maury et al.(2010)]{mau10} Maury, A.~J., et al.\ 2010, 
\aap, 512, A40 
\bibitem[Margulis \& Lada(1985)]{mar85} Margulis, M., \& Lada, C.~J.\ 
1985, \apj, 299, 925 
\bibitem[Moriarty-Schieven et al.(1987)]{mor87} 
Moriarty-Schieven, G.~H., Snell, R.~L., Strom, S.~E., Schloerb, F.~P., 
Strom, K.~M., \& Grasdalen, G.~L.\ 1987, \apj, 319, 742 
\bibitem[M{\"u}ller et al.(2001)]{mue01} M{\"u}ller, H.~S.~P., 
Thorwirth, S., Roth, D.~A., \& Winnewisser, G.\ 2001, \aap, 370, L49 
\bibitem[M{\"u}ller et al.(2005)]{mue05} M{\"u}ller, 
H.~S.~P., Schl{\"o}der, F., Stutzki, J., 
\& Winnewisser, G.\ 2005, Journal of Molecular Structure, 742, 215 
\bibitem[Neufeld \& Dalgarno(1989)]{neu89} Neufeld, D.~A., \& Dalgarno, 
A.\ 1989, \apj, 340, 869 
\bibitem[Nisini et al.(2007)]{nis07} Nisini, B., Codella, C., Giannini, T., 
Santiago-Garc\'{\i}a, J., Richer, J.~S., Bachiller, R., \& Tafalla, M.\ 
2007, \aap, 462, 163 
\bibitem[Panoglou et al.(2009)]{pan09} Panoglou, D., Cabrit, 
S., Garcia, P.~J.~V., 
\& For{\^e}ts, G.~P.~D.\ 2009, Protostellar Jets in Context, 459 
\bibitem[Pineau des For\^ets et al.(1990)]{pin90} Pineau des 
For\^ets, G., Roueff, E., \& Flower, D.~R.\ 1990, \mnras, 244, 668 
\bibitem[Pineau des For\^ets et al.(1993)]{pin93} Pineau des 
For\^ets, G., Roueff, E., Schilke, P., 
\& Flower, D.~R.\ 1993, \mnras, 262, 915 
\bibitem[Raga et al.(1990)]{rag90} Raga, A.~C., Binette, L.,
Canto, J., \& Calvet, N.\ 1990, \apj, 364, 601
\bibitem[Raga \& Cabrit(1993)]{rag93} Raga, A., \& Cabrit, S.\ 1993, \aap, 278, 267
\bibitem[Reipurth \& Bally(2001)]{rei01} Reipurth, B., \& Bally, J.\ 
2001, \araa, 39, 403 
\bibitem[Richer et al.(1992)]{ric92} Richer, J.~S., Hills, 
R.~E., \& Padman, R.\ 1992, \mnras, 254, 525 
\bibitem[Rodr\'{\i}guez-Fern\'andez et al.(2010)]{rod10} 
Rodr\'{\i}guez-Fern\'andez, N., Tafalla, M., Gueth, F., 
\& Bachiller, R.\ 2010, \aap, 516, A98
\bibitem[Santiago-Garc{\'{\i}}a et al.(2009)]{san09} Santiago-Garc{\'{\i}}a, 
J., Tafalla, M., Johnstone, D., \& Bachiller, R.\ 2009, \aap, 495, 169 
\bibitem[Schilke et al.(1997)]{sch97} Schilke, P., Walmsley, C.~M., 
Pineau des Forets, G., \& Flower, D.~R.\ 1997, \aap, 321, 293 
\bibitem[Shang et al.(2006)]{sha06} Shang, H., Allen, A., Li,
Z.-Y., Liu, C.-F., Chou, M.-Y., \& Anderson, J.\ 2006, \apj, 649, 845
\bibitem[Tafalla et al.(2004a)]{taf04} Tafalla, M., Myers, P.~C., 
Caselli, P., \& Walmsley, C.~M.\ 2004, \aap, 416, 191 
\bibitem[Tafalla et al.(2004b)]{taf04b} Tafalla, M., Santiago, J., 
Johnstone, D., \& Bachiller, R.\ 2004, \aap, 423, L21 
\bibitem[Tafalla et al.(2006)]{taf06} Tafalla, M., Santiago-Garc{\'{\i}}a, 
J., Myers, P.~C., Caselli, P., Walmsley, C.~M., \& Crapsi, A.\ 2006, 
\aap, 455, 577 
\bibitem[Tielens et al.(1994)]{tie94} Tielens, A.~G.~G.~M., 
McKee, C.~F., Seab, C.~G., \& Hollenbach, D.~J.\ 1994, \apj, 431, 321 
\bibitem[Tomida et al.(2010)]{tom10} Tomida, K., Tomisaka, 
K., Matsumoto, T., Ohsuga, K., Machida, M.~N., 
\& Saigo, K.\ 2010, \apjl, 714, L58 
\bibitem[Umemoto et al.(1992)]{ume92} Umemoto, T., Iwata, T., 
Fukui, Y., Mikami, H., Yamamoto, S., Kameya, O., 
\& Hirano, N.\ 1992, \apjl, 392, L83 
\bibitem[van Dishoeck \& Blake(1998)]{van98} van Dishoeck, E.~F., 
\& Blake, G.~A.\ 1998, \araa, 36, 317 
\bibitem[Zapata et al.(2010)]{zap10} Zapata, L.~A., Schmid-Burgk, J., 
Muders, D., Schilke, P., Menten, K., \& Guesten, R.\ 2010, \aap, 510, A2 
\bibitem[Ziurys et al.(1989)]{ziu89} Ziurys, L.~M., Friberg, 
P., \& Irvine, W.~M.\ 1989, \apj, 343, 201 
\bibitem[Watanabe et al.(2004)]{wat04} Watanabe, N., Nagaoka, 
A., Shiraki, T., \& Kouchi, A.\ 2004, \apj, 616, 638 








\end{thebibliography}
\end{document}